\documentclass[prd,preprint,tightenlines]{revtex4}

\usepackage{amsmath,amssymb}
\usepackage{bm} 

\usepackage{color} 
\usepackage{graphicx}

\usepackage[colorlinks]{hyperref} 
\hypersetup{
	citecolor=blue,
	linkcolor=blue,
	urlcolor=blue
}

\usepackage{epstopdf}
\RequirePackage[normalem]{ulem}
\usepackage{comment}

\usepackage{multirow}
\usepackage{threeparttable}
\usepackage{bm}
\usepackage{cleveref}

\usepackage{array}

\newcommand{\Sec}[1]{Sec.~\ref{#1}}  

\newcommand{\Fig}[1]{Fig.~\ref{#1}}

\newcommand{\Tab}[1]{Tab.~\ref{#1}}

\newcommand{\Eq}[1]{Eq.~(\ref{#1})}

\newcommand{\App}[1]{Appendix~\ref{#1}}  
 
\usepackage{mfirstuc} 
\renewcommand{\paragraph}[1]{} 
 
\newcommand{\beq}[1]{\begin{equation}\label{#1}}
\newcommand{\eeq}{\end{equation}}
 \newcommand{\bea}[1]{\begin{eqnarray}\label{#1}}
 \newcommand{\eea}{\end{eqnarray}}
 \newcommand\figcaption{\def\@captype{figure}\caption}
 \newcommand\tabcaption{\def\@captype{table}\caption}
\newcommand{\PYTHIA}{{\tt PYTHIA~8}}
\newcommand{\GALPROP}{{\tt GALPROP-v54}}

\newcolumntype{L}[1]{>{\raggedright\let\newline\\\arraybackslash\hspace{0pt}}m{#1}}
\newcolumntype{C}[1]{>{\centering\let\newline\\\arraybackslash\hspace{0pt}}m{#1}}
\newcolumntype{R}[1]{>{\raggedleft\let\newline\\\arraybackslash\hspace{0pt}}m{#1}}

\begin{document}

\title{Implications of a possible TeV break in the cosmic-ray electron and positron flux}
\author{
Yu-Chen Ding$^{1,2}$,
Nan Li$^{1,2}$, 
Chun-Cheng Wei$^{1,2}$,
Yue-Liang Wu$^{2,1,3,4}$, and 
Yu-Feng Zhou$^{1,2,3,4}$}
\affiliation{$^{1}$CAS Key Laboratory of Theoretical Physics, 
Institute of Theoretical Physics, Chinese Academy of Sciences, Beijing 100190, 
China.\\
$^{2}$University of Chinese Academy of Sciences, Beijing 100049, China.\\
$^{3}$School of Fundamental Physics and Mathematical Sciences, 
Hangzhou Institute for Advanced Study, UCAS, Hangzhou 310024, China.\\
$^{4}$International Centre for Theoretical Physics Asia-Pacific, Beijing/Hangzhou, China.}


\begin{abstract}
A TeV spectral break in the total flux of cosmic-ray electrons and 
positrons (CREs) at which the spectral power index softens from $\sim 3$
to $\sim 4$ has been observed by H.E.S.S. and recently confirmed by
DAMPE with a high significance of $6.6~\sigma$. Such an observation is 
apparently inconsistent with the data from other experiments such as 
Fermi-LAT, AMS-02, and CALET. We perform a global analysis to the latest CRE data
including Fermi-LAT, AMS-02, CALET, DAMPE, and H.E.S.S. with energy
scale uncertainties taken into account to improve the consistency
between the data sets. The fit result strongly favors the existence
of the break at $\sim 1$ TeV with an even higher statistical significance
of $13.3~\sigma$.  In view of the tentative CRE break, we revisit
a number of models of nearby sources, such as a single generic Pulsar
Wind Nebula (PWN), known multiple PWNe from the ATNF catalog, and
their combinations with either an additional Dark Matter (DM) component
or a Supernova Remnant (SNR).  We show that the CRE break at $\sim 1$ TeV,
together with the known CR positron excess 
points towards the possibility that the nearby sources should be highly
charge asymmetric. Among the models under consideration, the one
with a PWN plus SNR is most favored by the current data. The favoured
distance and age of the PWN and SNR sources are both within $0.6$ kpc
and around $10^{5}$ yr respectively. Possible candidate sources include
PSR J0954-5430, Vela and Monogem ring, etc. We find that for the models
under consideration, the additional DM component is either unnecessary
or predicts too much photons in tension with the H.E.S.S. data of $\gamma$-rays
from the direction of the Galactic Center. We also show that
the current measurement of the anisotropies in the arrival direction
of the CRE can be useful in determining the properties of the sources.
\end{abstract}

\maketitle

\section{Introduction}
High energy cosmic-ray electrons and positrons (CREs) lose their
energies rapidly via inverse Compton scattering and synchrotron radiation
during their propagation through the Galaxy. CREs with observed energy
above TeV typically come from a distance within 1 kpc, which
makes them an important probe of nearby sources. The existence of
the nearby sources may contribute to spectral features in the CRE
energy spectrum. Candidates of such CRE sources include: Pulsar Wind
Nebulae (PWNe), Supernova Remnants (SNRs), and Dark Matter (DM) particle
annihilation or decay, etc.

In recent years, with the successful running of several ground-
and space-based experiments, the measurement of the CRE flux in the
GeV--TeV region has been significantly improved. The CRE flux can be
measured by ground-based imaging atmospheric Cherenkov telescopes
such as H.E.S.S., VERITAS, and MAGIC. In 2009, the H.E.S.S. collaboration
measured CRE flux in the energy range 340 GeV -- 1.5 TeV,
and found a spectral break at $\sim 0.9$ TeV with the power-law index
softening from 3.0 to 4.1 \citep{Aharonian:2009ah}. 
The result was confirmed with higher statistics in the preliminary data of H.E.S.S. in 2018 \citep{HESSICRC17}.
The VERITAS collaboration observed a similar break at a lower significance
\citep{Archer:2018chh}. The results from MAGIC showed a trend of
spectrum softening after $\sim$TeV, but is in overall agreement with
a single power law in the energy range 100 GeV -- 3 TeV,
due to significant uncertainties \citep{BorlaTridon:2011dk}.

The satellite-borne Fermi-LAT experiment has measured the CRE spectrum
up to 2 TeV \citep{Abdollahi:2017nat}. Although the Fermi-LAT data
showed a trend of spectrum softening after around TeV, the whole CRE
spectrum above 47 GeV can still be well described by a single power
law with a power index $\sim 3.11$ due to significant energy reconstruction
uncertainties. The AMS-02 experiment onboard the international space
station (ISS) has measured the CRE spectrum up to 1 TeV (the CR electron
flux reached 1.4 TeV recently \citep{Aguilar:2019ksn}), but did not observe 
any significant structure below TeV \cite{Aguilar:2019ksn}. In 2017,
the satellite-borne experiment DAMPE which has relatively large acceptance
(compared with AMS-02) and high energy resolution (compared with Fermi-LAT)
has released the first measurement on the CRE spectrum up to 4.6 TeV,
which strongly favors a break at $\sim 0.9$ TeV at $6.6~\sigma$
significance. After the break, the CRE spectral power index softens
from $\sim 3.1$ to $\sim 3.9$ \citep{Ambrosi:2017wek}. Recently,
the CALET collaboration has extended the CRE measurement up to 4.8
TeV, and the result also supports a spectral softening at around 0.9
TeV \citep{Adriani:2018ktz}. The current CRE measurements from different
experiments are apparently not in full agreement with each other in
the energy range $\sim 30$ GeV -- a few TeV. Increasing
the statistics in the future is unlikely to resolve the problem as
the current data below TeV are dominated by systematic uncertainties.

In this work, we perform a global analysis of the latest CRE
data from Fermi-LAT, AMS-02, DAMPE, CALET, and H.E.S.S. We show that
a consistent fit of all the five data sets can be achieved by including
the uncertainties in the absolute energy scale of different experiments.
This part of uncertainty is usually not added to the total uncertainties
in the released data. The global fit result strongly favors the existence
of a spectral break at around TeV. After the break, the CRE spectral
power index softens from $\sim 3.10$ to $\sim 3.89$, which confirms
the result of DAMPE at a higher significance $\sim 13.3~\sigma$.

In view of the tentative CRE break, we revisit a number of models
of nearby sources, such as a single generic Pulsar Wind Nebula (PWN),
known multiple PWNe from the ATNF catalog, and their combinations
with either an additional dark matter (DM) component or a Supernova
Remnant (SNR). In total six models are considered: A) a single PWN,
B) all middle-aged PWNe from the ATNF catalog
with a simplified assumption that they share a common spectral index and efficiency,
C) a single PWN plus
an additional DM component which annihilates directly into $2\mu$ final states,
D) all middle-aged PWNe plus an additional DM component which annihilates
directly into $2\mu$ final states, E) a single PWN plus a single
SNR, F) middle-aged PWNe plus a single SNR and DM. The CRE break at
$\sim \text{TeV}$, together with the CR positron spectrum
peaking at $\sim 300$ GeV suggests the possibility
that the nearby sources would be highly charge asymmetric. Consequently,
we find that among these models, only models E and F can well account
for the current CRE and CR positron spectra simultaneously. For model-E,
the data favor a nearby middle-aged PWN with a spectral index $\sim2$
and an energy cutoff at $\sim0.8$ TeV. PSR J0954-5430 is a possible PWN
candidate. The favored additional SNR turns out
to have a spectral index $\sim2.2$ and a total energy $\sim5.5\times10^{48}$
erg. Possible SNR candidates include Vela and Monogem ring. For model-F,
the data favor a DM particle with mass $\sim$ 1 TeV and annihilation
cross-section $\sim1.69\times10^{-24}~{\rm cm^{3}s^{-1}}$. The favored DM
parameters are consistent with the limits derived from Fermi-LAT data
of $\gamma$-rays from dwarf galaxies \citep{Ackermann:2015zua},
but still in tension with the H.E.S.S. data from the Galactic Center
(GC) \citep{Abdallah:2016ygi}. 
In model-F, the middle-aged PWNe turn out to have a spectral $\sim2$, 
and an efficiency $\sim0.098$.
The favored additional SNR for this model turns out to have a
spectral index $\sim1.9$ and a total energy $\sim4.1\times10^{48}$
erg. Possible SNR candidates include Vela and Monogem ring. 
In addition, we predict the dipole anisotropy in
CRE flux for relevant models and compare them with the current upper-limits
obtained by Fermi-LAT \citep{Abdollahi:2017kyf}. We find that the upper-limits
on the CRE anisotropy could be a useful tool for understanding the
properties of the $e^{\pm}$ sources in these models.

This paper is organized as follows: In \Sec{sec:epproduce}, we briefly
overview the calculation of CR propagation and possible sources of
CREs. In \Sec{sec:Experimental_data}, we perform a global analysis
of the current CRE data. In \Sec{sec:ModelAndResults}, we discuss
a number of models of nearby sources. 
In \Sec{sec:discuss}, we summarize the conclusions of this work.
The impact of the uncertainties in the propagation models on the 
conclusion is also discussed.

\section{PROPAGATION OF COSMIC RAYS IN THE GALAXY\label{sec:epproduce}}

\subsection{The propagation model \label{sec:propagation_model} }

The propagation of CR particles through the Galaxy can be approximated
by a diffusion model in which the diffusion halo is parameterized
by a cylinder with radius $R\simeq20$ kpc and half-height $Z_{h}=1\sim10$
kpc. The diffusion equation for the CR charged particles reads \citep{Ginzburg:1990sk,Strong:2007nh}
\begin{equation}
\frac{\partial\psi}{\partial t}=\nabla(D_{xx}\nabla\psi-\boldsymbol{V}_{c}\psi)+\frac{\partial}{\partial p}p^{2}D_{pp}\frac{\partial}{\partial p}\frac{1}{p^{2}}\psi-\frac{\partial}{\partial p}\left[\frac{\mathrm{d}p}{\mathrm{d}t}\psi-\frac{p}{3}(\nabla\cdot\boldsymbol{V}_{c})\psi\right]-\frac{1}{\tau_{f}}\psi-\frac{1}{\tau_{r}}\psi+Q(\boldsymbol{r},t,p),\label{eq:propagation}
\end{equation}
where $\psi(\boldsymbol{r},t,p)$ is the CR number density per unit
momentum, $D_{xx}$ is the spatial diffusion coefficient, and $\boldsymbol{V}_{c}$
is the convection velocity. The re-acceleration effect is described
as diffusion in momentum space and is determined by the coefficient
$D_{pp}$. The quantity $dp/dt$ stands for the momentum loss rate.
$\tau_{f}$ and $\tau_{r}$ are the time scales for fragmentation
and radioactive decay respectively. 
$Q(\boldsymbol{r},t,p)$ is the source term.
The energy-dependent spatial diffusion coefficient
$D_{xx}$ is parameterized as $D_{xx}=\beta D_{0}\left({\rho}/{\rho_{0}}\right)^{\delta}$,
where $\rho=p/Ze$ is the rigidity of CR particles with electric charge
$Ze$,
$\delta$ is the spectral power index, $\rho_{0}$ is a reference
rigidity,
$D_{0}$ is a normalization constant, and $\beta=v/c$
is the velocity of CR particles. The momentum diffusion coefficient
$D_{pp}$ is related to $D_{xx}$ as $D_{pp}D_{xx}=4V_{a}^{2}p^{2}/(3\delta(4-\delta^{2})(4-\delta))$,
where $V_{a}$ is the Alfv\`{e}n velocity of disturbances in the hydrodynamical
plasma \citep{Ginzburg:1990sk}. 
The source term of primary CR particles
is expressed as $Q(\boldsymbol{r},t,p)=f(\boldsymbol{r},t)q(p)$,
where $f(\boldsymbol{r},t)$ is the spatial distribution
and $q(p)$ is the injection spectrum.
The spatial distribution of the source is taken from Ref. \citep{1996A&AS..120C.437C}. 
The injection spectra of the primary nucleus are assumed to be 
a broken power law behavior, 
$q(p)\propto({\rho}/{\rho_{s}})^{\gamma_{\rm nucl}}$,
with the injection index $\gamma_{\rm nucl}=\gamma_{\rm nucl,1}(\gamma_{\rm nucl,2})$ for
the nucleus rigidity $\rho$ below (above) a reference rigidity $\rho_{s}$.
The spatial boundary conditions are set by assuming that free particles
escape beyond the halo, i.e., $\psi(R,z,p)=\psi(\boldsymbol{r},\pm Z_{h},p)=0$.
The steady-state solution can be obtained by setting $\partial\psi/\partial t=0$. 

In this work, we use the public code \GALPROP~\citep{Strong:1998pw,MOSKALENKO:2001YA,STRONG:2001FU,MOSKALENKO:2002YX,PTUSKIN:2005AX}
to numerically solve this equation for the CR propagation. The propagation
parameters are fixed to the ``MED'' diffusion re-acceleration (DR)
propagation model, which is obtained from a global fit to the proton
and B/C data of AMS-02 using the GALPROP code \citep{Jin:2014ica}.
Note that this model is different from the one proposed in Ref. \citep{Donato:2003xg}
which is based on semi-analytical solutions of the propagation equation. 
The main parameters
of this model are summarized in \Tab{tab:med_model}. When CR particles
propagate into the heliosphere, the flux of CR particles is 
affected by the solar wind and the heliospheric magnetic field.
In order to account for the solar modulation, we adopt the force field 
approximation with a modulation potential $\phi=0.55$, which is 
consistent with the value adopted in deriving the ``MED'' model 
from the experimental data.

\begin{table}[t]
  \begin{tabular}{*{7}{C{1.8 cm}} C{2.2 cm}}\hline\hline
    $R(\mbox{kpc})$ & $Z_{h}(\mbox{kpc})$ & $D_{0}$ & $\rho_{0}$(GV) & $\delta$ & $V_{a}(\mbox{km}/\mbox{s})$ & $\rho_{s}$(GV) & $\gamma_{\rm nucl,1}/\gamma_{\rm nucl,2}$ \\ \hline
    20 & 3.2 & 6.50 &4.0 & 0.29 & 44.8 & 10.0 & 1.79/2.45 \\ \hline\hline
  \end{tabular}
  \caption{Values of the main parameters in the ``MED'' propagation model 
  derived from fitting to the AMS-02 B/C and proton data based on 
  the GALPROP code \cite{Jin:2014ica}. The parameter $D_{0}$ is in units 
  of $10^{28}~\mbox{cm}^{2}\cdot\mbox{s}^{-1}$.}
  \label{tab:med_model}
\end{table}

\subsection{Sources of primary and secondary CRE}
SNRs in our Galaxy are often considered as the major source of primary
CR particles. 
Charged particles can be accelerated to a very high energy by
non-relativistic diffusive shock wave through the Fermi acceleration 
mechanism \cite{1977DoSSR.234.1306K,Blandford:1978ky,Bell:1978zc,Bell:1978fj}.
The supernova explosion rate in the Galaxy is $\sim3$ per century.
Thus the injection of the primary CR particles from the SNRs can be
assumed to be a stable continuous source. The source term of the primary
electrons from the SNRs can be written as 
\begin{equation}
Q_{{\rm pri}}(\boldsymbol{r},t,p)=f(r,z)q_{{\rm pri}}(p),
\end{equation}
where $f(r,z)$ is the spatial distribution and $q_{{\rm pri}}(p)$
is the injection spectrum of the source. The spatial distribution
is assumed to follow the SNRs distribution \citep{1996A&AS..120C.437C}
\begin{equation}
f(r,z)=\left(\frac{r}{r_{\odot}}\right)^{a}\exp\left(-b\cdot\frac{r-r_{\odot}}{r_{\odot}}\right)\exp\left(-\frac{|z|}{z_{s}}\right),\label{eq:frz}
\end{equation}
where $r_{\odot}=8.5$ kpc is the distance from the Sun to the GC, 
$z_{s}\approx0.2$ kpc is the characteristic height of the
Galactic disk. The two parameters $a$ and $b$ are chosen to be $a=1.25$
and $b=3.56$, which are adopted to reproduce the Fermi-LAT $\gamma$-ray gradient
\citep{Trotta:2010mx,Tibaldo:2009spa}. The typical injected
spectra have the shape of a power-law with an exponential cut-off:
\begin{equation}
q_{{\rm pri}}(\rho)\propto\left(\frac{\rho}{1{\rm GV}}\right)^{-\gamma_{e}}\exp\left(-\frac{\rho}{\rho_{c}}\right),\label{eq:primary_e_sp}
\end{equation}
where $\gamma_{e}$ is the power index, and $\rho_{c}$ is the exponential
cutoff in rigidity.

During the propagation, the spallation process of the CR nuclei in
the interstellar medium (ISM) will produce secondary particles. The
corresponding source term is given by 
\begin{equation} \label{eq:sec_e}
Q_{{\rm sec}}(p)=\sum_{i=\mathrm{H,He}}n_{i}\sum_{j}\int c\beta_{j}n_{j}(p')\frac{d\sigma_{ij}(p,p')}{dp}dp'
\end{equation}
where $n_{i}$ is the number density of the interstellar gas, $n_{j}(p')$
is the number density of CR particles, $d\sigma_{ij}(E,p')/dE$ is
the differential cross-section for the production of electrons and
positrons from the interaction between CR particles and the interstellar
gas. 
The production cross-sections can be obtained through parameterizations 
of the available \emph{pp}-collision data \cite{Kamae:2006bf,
Tan:1984ha,Badhwar:1977zf,1986ApJ...307...47D,1986A&A...157..223D}, 
or using Monte-Carlo event generators with QCD inspired phenomenological models
\cite{Huang:2006bp,Sjostrand:2006za}.
The choice of different cross-section parameterizations can result in the change of the 
secondary electron/positron flux up to 30\% \cite{Feng:2016loc,Evoli:2017vim,
Lipari:2016vqk}. The change in the cross-sections only leads to minor changes
in the analysis, since the secondary electrons and positrons are subdominant
in the high energy region. In this work, we employ the parameterization 
in \cite{1986ApJ...307...47D,1986A&A...157..223D}
which is implemented in \GALPROP~\cite{Moskalenko:1997gh}.
As it can be seen from \Eq{eq:sec_e}, for a given propagation 
model, the fluxes of the secondary CR electrons and positrons can be 
predicted from the distribution of primaries without free parameters.

In the GALPROP code, the primary electron source term is normalized
in such a way that the flux of the primary electrons at a reference
kinetic energy $E_{{\rm ref}}$ is reproduced. 
In this work, we fix the value of $E_{{\rm ref}}$ at $E_{{\rm ref}}=25$
GeV and fit the post-propagated normalization flux $N_{e}$ to the
data. 


\subsection{CRE anisotropy}\label{sec:anisotropy} 
CREs at high energies are most probably from
nearby sources, which may lead to visible anisotropy of CRE flux in
the arrival direction. The measurement on CRE anisotropy can be a
useful tool to constrain the properties of nearby CR source, which
is complementary to the data of the total flux. In the diffusion model,
the dipole anisotropy of CRE flux is given by
\begin{equation}
\Delta(E)=\frac{3D_{xx}(E)}{c}\frac{|\nabla\psi(E)|}{\psi(E)},\label{aniso2}
\end{equation}
where $\psi(E)$ is the number density of CRE, and $c$ is the speed
of light. For a collection of CRE sources, the total dipole anisotropy
can be computed as \citep{1971ApL.....9..169S}: 
\begin{equation}
\Delta\left(n_{\max},E\right)=\frac{1}{\psi_{\operatorname{tot}}(E)}\cdot\sum_{i}\frac{\mathbf{r}_{i}\cdot\mathbf{n}_{\max}}{\left\Vert \mathbf{r}_{i}\right\Vert }\cdot\psi_{i}(E)\Delta_{i}(E),\label{aniso_sum}
\end{equation}
where $\psi_{i}(E)$ is the number density of CRE from each source
$i$, $\mathbf{r}_{i}$ is the position of the source, $\Delta_{i}(E)$
is the dipole anisotropy from each source from
\Eq{aniso2}, $\mathbf{n}_{\max}$ is the direction of maximum flux
intensity, and $\psi_{\operatorname{tot}}(E)=\sum_{i}\psi_{i}(E)$ is the 
total CRE number density. 

\section{a global analysis of the current CRE data \label{sec:Experimental_data}}

All the current measurements from different experiments show 
that the energy spectrum of CRE flux approximately follows a power
law in a wide energy range, however, different experiments are apparently not
in full agreement in details: $i)$ for energies below $\sim$ 140
GeV, the DAMPE data is slightly higher than that from all the other
experiments. Starting from $\sim$ 70 GeV, the Fermi-LAT data of the CRE spectrum is harder
than that measured by AMS-02 and CALET; $ii)$ for the energy range
140 GeV -- 1 TeV, the data of Fermi-LAT and DAMPE is noticeably higher
than that from AMS-02 and CALET while the H.E.S.S. data in this energy
range is compatible with that from in AMS-02 and CALET; $iii)$ for
energies above $\sim$ 1 TeV, the measured CRE spectra of DAMPE, CALET and
H.E.S.S. all start to soften, while no significant spectral change
was observed by Fermi-LAT in this energy region. 
Increasing
the statistics in the future is unlikely to resolve the problem as
the current data below TeV are dominated by systematic uncertainties.
The absolute energy scale calibration is one of the key parts in 
the CRE flux measurements. 
The absolute energy scale is defined as $S\equiv E/E_{t}$,
where $E_{t}$ and $E$ are the true and measured energies respectively.
The measured flux $\Phi(E)$ is related to the true
flux $\Phi_{t}(E_{t})$ by $\Phi(E)=\Phi_{t}\left(E_{t}\right)/S$.
The value of $S$ and its relative uncertainty $\delta_{s}=\Delta S/S$
can be determined by using the geomagnetic cutoff at around 10 GeV. 
The released experimental data usually include the corrections for 
the absolute energy scale $S$, while the uncertainties in $S$ (i.e. $\delta_{s}$)
are not included. The consistency between the 
current CRE data can be improved by considering the uncertainties
in the absolute energy scale. A full treatment of energy
scale uncertainty in each experiment is complicated. In this work, we adopt a naive
method of the error propagation in which the systematic uncertainty on the CR flux
due to the energy scale uncertainty is estimated by 
\begin{equation}
\Delta\Phi/\Phi\approx\left|\frac{E}{\Phi(E)}\frac{d\Phi}{dE}+1\right|\delta_{s},\label{eq:flux_uncertainty}
\end{equation}
For a signal power-law spectrum with a spectral index $\gamma$, the
above expression can be simplified to $\Delta\Phi/\Phi\approx|\gamma-1|\delta_{s}$,
which is in agreement with the literature \citep{Ackermann:2010ij}.
The details of the estimation of $\Phi(E)$, $d\Phi/dE$ and $\Delta\Phi/\Phi$
for each experimental data set are given in \App{app:flux_correction}.

The values of the energy scale uncertainty for each experiment
are summarized below:
\begin{itemize}
\item AMS-02 (both CRE and CR positron measurement), the typical value of
$\delta_{s}$ is $\sim4\%$ at 0.5 GeV, $\sim2\%$ from 2 GeV to 300
GeV, and $\sim2.6\%$ at 1.4 TeV \citep{Aguilar:2019ksn,Aguilar:2019owu,Kounine:2017pss}; 
\item Fermi-LAT, $2\%$ for the whole energy range \citep{Abdollahi:2017nat}; 
\item DAMPE, the current preliminary value is $1.26\%$ in the whole energy
range \citep{DAMPEISM19}; 
\item CALET, the energy scale is determined by two independent
methods, geomagnetic cutoff and MIP calibrations \citep{Adriani:2017efm,Adriani:2018ktz}.
The corresponding values are $1.035\pm0.009~(\text{stat})$ and $1.000\pm0.013~(\mathrm{sys})$,
respectively. The reason for such a difference is yet to be understood.
Since the released CALET data already include the correction of the
absolute energy scale $S$ (not the uncertainty $\delta_{s}$) using
the value from the geomagnetic cutoff \citep{Adriani:2018ktz}, we
take the corresponding uncertainty to be $\delta_{s}=0.009$. 
\item H.E.S.S. the value is $15\%$ which is much larger than that from
the space-based experiments \citep{Aharonian:2009ah}. 
\end{itemize}

\begin{table}[t]
  \begin{center}
  \scalebox{0.88}{
  \begin{threeparttable}
  \begin{tabular}{L{2.1cm} C{2.3cm} C{2.1cm} C{2.1cm} C{2.2cm} C{1.9cm} C{2.7cm} | C{2.0cm} }    \hline \hline
      & $\Phi_{0}$ & $\gamma_{1}$ & $\gamma_{2}$ & $\gamma_{3}$ & $E_{\mathrm{br1}}$ & $E_{\mathrm{br2}}$ & $\bm{\chi^{2} / \mathrm{d.o.f.}}$ \\ \hline
  FERMI\tnote{$\dagger$}     & 5.39 $\pm$ 0.05 & 3.25 $\pm$ 0.02   & 3.06 $\pm$ 0.01   & 3.27 $\pm$ 0.13   & 49.7 $\pm$ 3.4    & 847.7 $\pm$ 293.8 & $\bm{2.8/31}$ \\
  DAMPE                      & 5.42 $\pm$ 0.05 & 3.20 $\pm$ 0.08   & 3.09 $\pm$ 0.01   & 4.01 $\pm$ 0.18   & 45.9 $\pm$ 16.5   & 925.2 $\pm$ 92.4  & $\bm{25.6/32}$ \\
  CALET                      & 4.57 $\pm$ 0.05 & -                 & 3.15 $\pm$ 0.01   & 3.83 $\pm$ 0.33   & -                 & 959.5 $\pm$ 225.5 & $\bm{13.6/29}$ \\
  AMS-02                     & 4.68 $\pm$ 0.03 & 3.24 $\pm$ 0.02   & 3.13 $\pm$ 0.00   & -                 & 45.4 $\pm$ 4.3    & -                 & $\bm{9.4/31}$ \\
  H.E.S.S.                   & 4.04 $\pm$ 0.12 & -                 & 3.02 $\pm$ 0.07   & 3.67 $\pm$ 0.02   & -                 & 804.8 $\pm$ 78.5  & $\bm{0.2/14}$ \\ \hline
  \textbf{Global Fit}          & 5.00 $\pm$ 0.05 & 3.23 $\pm$ 0.03   & 3.10 $\pm$ 0.01   & 3.86 $\pm$ 0.16   & 52.6 $\pm$ 7.6    & 975.9 $\pm$ 131.3 & $\bm{302.3/155}$ \\
  \textbf{Global Fit}\tnote{*} & 5.02 $\pm$ 0.05 & 3.24 $\pm$ 0.07   & 3.10 $\pm$ 0.01   & 3.89 $\pm$ 0.14   & 46.0 $\pm$ 11.2   & 987.8 $\pm$ 110.2 & $\bm{170.5/155}$ \\
  \hline\hline
  \end{tabular}
  \begin{tablenotes}
  \footnotesize
  \item[$\dagger$] the LAT energy reconstruction uncertainties are considered.
  \item[*] the absolute energy scale uncertainties are considered.
  \end{tablenotes}
  \end{threeparttable}}
  \end{center}
  \caption{The fit parameters corresponding to the fit of \Eq{eq:smooth_pl} to 
  the CRE data with energies above 25 GeV from Fermi-LAT \cite{Abdollahi:2017nat}, 
  DAMPE \cite{Ambrosi:2017wek}, CALET \cite{Adriani:2018ktz}, AMS-02 \cite{Aguilar:2019ksn},
  H.E.S.S. \cite{HESSICRC17} and to the CRE data from all the five experiments
  with and without including the energy scale uncertainties. 
  The reduced $\chi^2$ of each fit is also listed. $\Phi_{0}$ is in units of 
  $10^{-6}~\mathrm{m}^{-2} \mathrm{sr}^{-1} \mathrm{s}^{-1} \mathrm{GeV}^{-1}$,
  $E_{\mathrm{br1}}$ and $E_{\mathrm{br2}}$ are in units of $\mathrm{GeV}$.}
  \label{tab:spl_fiteach25GeV}
  \end{table}

We perform a global fit to the CRE flux data with a smoothly broken
power-law \citep{Ackermann:2013wqa,Ambrosi:2017wek}
\begin{equation}
\Phi_{\mathrm{CRE}}(E)=\Phi_{0}\left[1+\left(\frac{E_{\mathrm{br1}}}{E}\right)^{k}\right]^{(\gamma_{1}-\gamma_{2})/k}\left(\frac{E}{300\mathrm{GeV}}\right)^{-\gamma_{2}}\left[1+\left(\frac{E}{E_{\mathrm{br2}}}\right)^{k}\right]^{-(\gamma_{3}-\gamma_{2})/k},\label{eq:smooth_pl}
\end{equation}
where $\Phi_{0}$ is a normalization factor, $E_{\mathrm{br1,br2}}$
are the break energies and $\gamma_{1,2,3}$ are the spectral power
indexes. The smoothness parameter is fixed to $k=10$. Other parameters
are determined through maximizing the log likelihood (or minimizing the
$\chi^{2}$ function)
\begin{equation}
-2\ln\mathcal{L} = \chi^{2}=\sum_{i}\frac{\left(\Phi_{i}-\Phi_{\exp,i}\right)^{2}}{\sigma_{\exp,i}^{2}},
\end{equation}
where $\Phi_{i}$ is the theoretical value, $\Phi_{\exp,i}$ and $\sigma_{\exp,i}$
are the measured value and uncertainty of the CRE flux. We consider
the data in the energy range 25 GeV $\sim$ 15 TeV. For the Fermi-LAT
CRE data, the energy reconstruction uncertainty is included. The low
energy break $E_{\mathrm{br1}}$ is not used for the fit to the CALET
data, as it can not be constrained by the CALET data for energies
above 25 GeV. 

We first fit to the CRE data of each experiment independently, and
then make a global fit to the data from all the five experiments with
and without including the energy scale uncertainties. The best-fit
parameters and the goodness-of-fit are summarized in \Tab{tab:spl_fiteach25GeV}.

\begin{figure}[t]
  \centering
  \begin{center}
    \includegraphics[width=0.7\columnwidth]{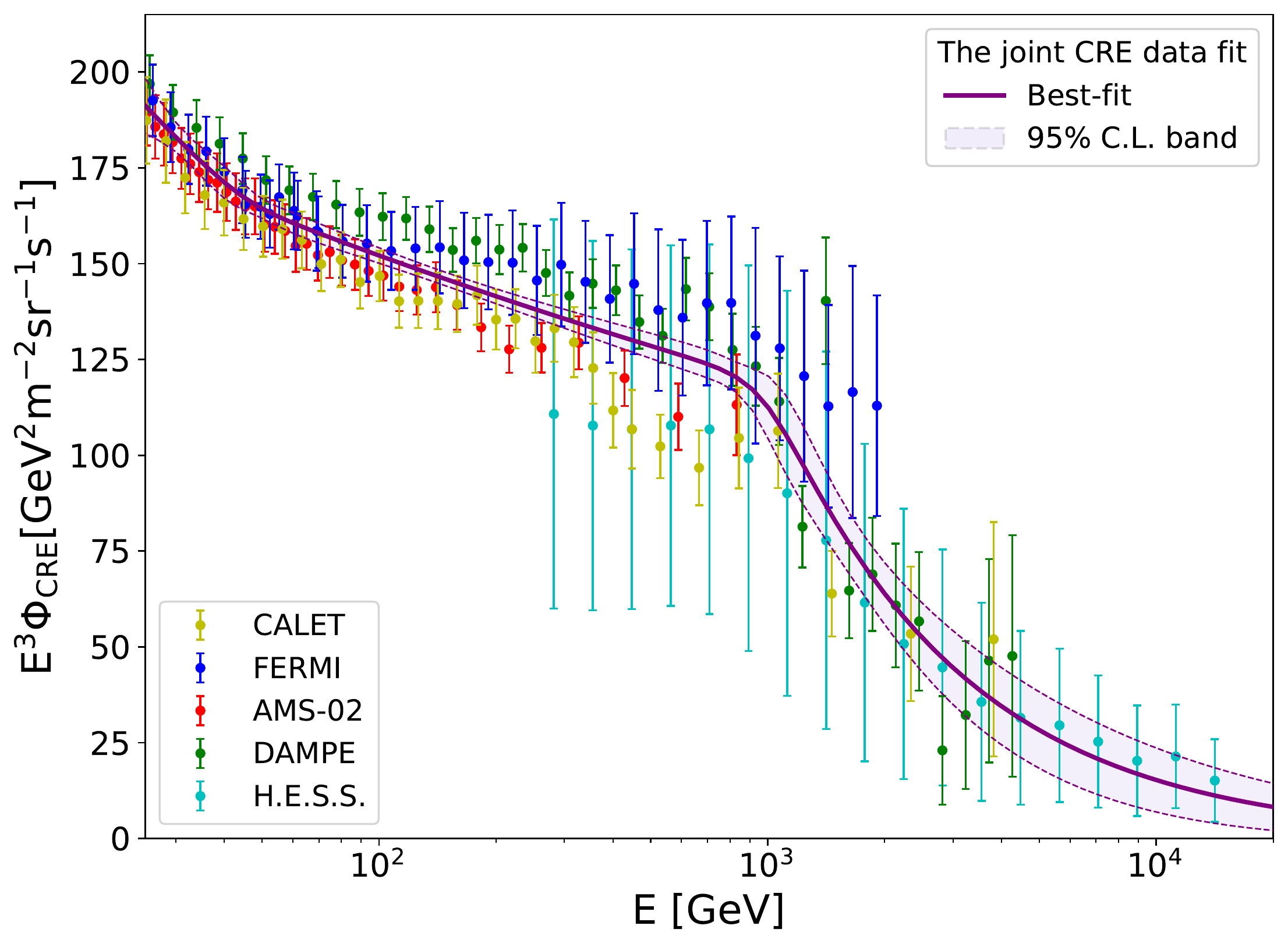}
  \end{center}
  \caption{CRE flux from a global fit to the latest CRE data from all five experiments
  (Fermi-LAT \cite{Abdollahi:2017nat}, DAMPE \cite{Ambrosi:2017wek}, 
  CALET \cite{Adriani:2018ktz}, AMS-02 \cite{Aguilar:2019ksn}, and H.E.S.S. 
  \cite{HESSICRC17}) with including the energy scale uncertainties. The purple 
  band corresponds to the parameters varying within 95\% C.L. The error 
  bars represent the total uncertainties (quadratic sum of the 
  statistical and systematic uncertainties) with the energy scale uncertainties
  included.}
  \label{fig:totalCREfit}
\end{figure}

From \Tab{tab:spl_fiteach25GeV}, one can see that without including
the energy scale uncertainty the global fit leads to $\chi^{2}/\mathrm{d.o.f.}\approx302.3/155$.
After including this part of uncertainty, the fit result gives $\chi^{2}/\mathrm{d.o.f.}\approx170.5/155$,
which is a significant improvement of the goodness-of-fit. Thus the
consistency between the data sets is improved after including the
energy scale uncertainty. In \Fig{fig:totalCREfit}, we show the
best-fit CRE flux from the global fit together with its 95\% C.L.
uncertainty band.

The individual fits to each experiment show that most current measurements
(expect for AMS-02) favor the existence of a spectral break at around
TeV at some extent. To quantify the significance of the spectral break, we adopt
a test statistic of $\text{TS}=-2\ln(\mathcal{L}_{0}/\mathcal{L})$,
where $\mathcal{L}(\mathcal{L}_{0})$ stands for the likelihood with
(without) the break. We find that even the Fermi-LAT data slightly
favor a high energy spectral break $E_{\mathrm{br2}}$ at $847.7\pm293.8$
GeV with $\chi^{2}/\mathrm{d.o.f.}=2.8/31$. Another fit without the
energy break $E_{\mathrm{br2}}$, results in $\chi^{2}/\mathrm{d.o.f.}=3.6/33$.
Thus roughly corresponding to a significance of $\sim0.43~\sigma$
for a $\chi^{2}$-distribution with two degrees of freedom, which
is consistent with the conclusion of the Fermi-LAT collaboration,
i.e., not observing significant structures in the high energy range.

The significance of the high energy spectral break in the global fit
is estimated as follows: We perform global fits in the energy range
of 55 GeV -- 15 TeV for two models. One is a single power-law
model $\Phi=\Phi_{0}(E/300~\mathrm{GeV})^{-\gamma}$ and the other
one is a smoothly broken power-law model $\Phi=\Phi_{0}(E/300~\mathrm{GeV})^{-\gamma_{1}}[1+(E/E_{\mathrm{b}})^{k}]^{-(\gamma_{2}-\gamma_{1})/k}$
with $k=10$. The single power-law fit gives $\gamma=3.16\pm0.01$,
with $\chi^{2}/\mathrm{d.o.f.}=338.8/122$. The smoothly broken power-law
fit gives $\gamma_{1}=3.10\pm0.01$, $\gamma_{2}=3.89\pm0.16$, $E_{\mathrm{b}}=988\pm120$,
and $\chi^{2}/\mathrm{d.o.f.}=157.0/120$. Compared to the single
power-law model, the $\chi^{2}$ value is reduced by 181.8 for two
less degrees of freedom. From the TS value we estimate the significance
of the spectral break is $\sim13.3~\sigma$. The significance from
the global fit is even higher than that from the DAMPE data alone,
which is easy to understand as the other experiments (except for AMS-02)
also favor the existence of the spectral break.

\section{Theoretical Models and Results \label{sec:ModelAndResults} }

The TeV break in the CRE flux, if confirmed, may constitute another
CR lepton anomaly complementary to the well-known CR positron excess.
The break may originate from extra sources or non-standard mechanisms
of CR acceleration and propagation. In this work, we shall focus on
the first probability. Possible nearby extra sources include PWNe,
SNRs and DM annihilation or decay, etc. We investigate what kind
of sources or combination of sources can account for \emph{both} the
CRE break and the CR positron excess. 

\subsection{Charge-symmetric electron–positron sources \label{sec:c_sym_sources} }

We first consider charge-symmetric sources which produce equal amount
of CR electrons and positrons with the same energy spectrum. PWNe
and DM annihilation are among such kind of sources.


PWNe are well-known powerful sources of primary electrons and positrons
which are believed to be produced in the magnetosphere and accelerated 
by the termination shock. The time-evolution of the luminosity of 
the associated pulsar can be written as 
$\dot \xi = \dot \xi_0 \left(1 + t/\tau_0\right)^{-\frac{n+1}{n-1}}$,
where $\dot \xi_0$ is the initial spin-down luminosity, 
$t$ is the age of the  pulsar, $\tau_0$ is the luminosity decay time, and 
$n$ is the breaking index \cite{Gaensler:2006ua}. 
This expression can be derived from a largely model-independent assumption 
where the pulsar spin-down is described by $\dot{\Omega}=-K \Omega^{n}$, 
where $\Omega$ is the rotation frequency and $K$ is a global constant. 
If $t/\tau_0 \gg 1$, the luminosity drops rapidly and can be considered as 
a burst-like source of energy release. 
In the opposite limit where $t/\tau_0 \ll 1$,
the luminosity can be approximated as a constant in time. 

In this work, we shall simply model the emission of  electrons/positrons from PWNe 
as a burst-like process, as commonly adopted in the literature \cite{Hooper:2008kg,
Yin:2013vaa,Wang:2017hsu,Malyshev:2009tw,Grasso:2009ma,Delahaye:2010ji,
DiBernardo:2010is,DiMauro:2014iia,Manconi:2016byt,DiMauro:2017jpu,Manconi:2018azw}. 
In the burst-like scenario,
the source term of a PWN at origin and $t=0$ is assumed to have a power-law
energy spectrum with an exponential cutoff
\begin{equation}
Q(E,\boldsymbol{r},t)=Q_{0}\left(\frac{E}{{\rm GeV}}\right)^{-\gamma}\exp\left(-\frac{E}{E_{c}}\right)\delta(\boldsymbol{r})\delta(t),\label{eq:burst_like_source}
\end{equation}
where $\gamma$ is the spectral power index and $E_{c}$ is the cutoff
energy. The normalization factor $Q_{0}$ is related to the total
injected energy $E_{\mathrm{tot}}$ by
\begin{equation}
E_{\mathrm{tot}}=\int_{E_{\text{min}}}^{\infty}EQ(E,\boldsymbol{r},t)dEd\boldsymbol{r}dt
\end{equation}
where the integration lower limit is set to $E_{\text{min}}=0.1$
GeV. For PWNe the total injection energy is assumed to be a fraction
$\eta$ of the total spin-down energy $W_{0}$ of the associated pulsar,
namely $E_{\text{tot}}=\eta W_{0}$.  
In the magnetic-dipole (MD) emission model
the value of $W_0$ is given by $W_{0}=\dot{\mathcal{E}}t(1+t/{\tau_{0}})$, 
where $t$ and $\dot{\mathcal{E}}$ are the age and the spin-down luminosity
of the pulsar, respectively. 
The typical luminosity decay time $\tau_{0}$ is taken
to be $10$ kyr~\cite{Aharonian:1995zz}, 
which is commonly adopted in the literature (see e.g. \cite{Hooper:2017gtd,
Yin:2013vaa,Wang:2017hsu,DiMauro:2014iia,Manconi:2016byt}).
%
%
%
Note that a constant emission of electrons from PWN is also possible and has been extensively discussed. 
For instance, in a recent analysis \cite{Fornieri:2019ddi}
the authors approximated pulsar age $t$ with its characteristic age
$t_{\rm ch} = P/2\dot P$, 
and calculated the luminosity decay time using the MD model in which
$\tau_{0}^{\mathrm{MD}}=3 I c^{3}/B_p^{2} R^{6} \Omega_{0}^{2}$, 
where $I$ is the moment of inertia, $B_p$ is the polar surface magnetic field, and
$\Omega_0$ is the initial frequency. 
They estimated the ratio $t_{\rm ch}/\tau_0^{\rm MD}$ for the nearby 
pulsars in the ATNF catalog and found that the typical  values of $t/\tau_0$ are  lower than unity ($\sim 0.3$), which points towards a constant energy injection.
Note, however, that so far a reliable estimation of the initial period ($P_0 = 2\pi/\Omega_0$)  is only available for a very limited number of pulsars \cite{2004hpa..book.....L,FaucherGiguere:2005ny}, 
which requires  independent estimation of the true age $t$ of the related pulsar 
(e.g. the pulsar may have  a historical association of a known supernova explosion to determine its true age $t$),  and precise measurement of the pulsar braking index. We have estimated the ratio $t/\tau_0^{\rm MD}$ for the pulsars 
which have proper estimation of $\Omega_0$  summarized in Table 7 of~\cite{FaucherGiguere:2005ny}, and find that the values  are in a wide range $\sim 0.1 - 2.1$, which suggests  that the characteristic of luminosity evolution 
with time for different pulsars could be different. 
%
Note also that the pulsars associated with constant-luminosity injection are typically young with  ages around a few thousand years. The ones with older ages are more likely to be associated with the burst-like injection. 
For more detailed discussions on the electron/positron emission time scale in PWNe, we refer to  \cite{Fornieri:2019ddi} and references therein. 


For high energy CR electrons/positrons, the propagation can be simplified by neglecting
convection and re-acceleration as they are only important at low energies.
Thus only the energy-dependent diffusion and the energy loss due to
synchrotron and Inverse Compton scatterings are relevant. The energy-dependent
spatial diffusion coefficient is assumed to be $D(E)\simeq D_{0}\left({E}/4{\rm GeV}\right)^{\delta}$.
The energy-loss rate is parameterized as ${\mathrm{d}E}/{\mathrm{d}t}=-b_{0}E^{2}$
with $b_{0}=1.4\times10^{-16}~{\rm GeV^{-1}s^{-1}}$ \cite{Grasso:2009ma}. 
For nearby sources,
one can adopt a spherically symmetric boundary condition such that
the Green's function of the propagation equation for the burst-like source
in \Eq{eq:burst_like_source} can be obtained analytically
\citep{Atoian:1995ux}
\begin{equation}
  \psi(E,t,\boldsymbol{r})=\frac{Q_{0}}{\pi^{3/2}r_{{\rm diff}}^{3}}\left(1-\frac{E}{E_{{\rm max}}}\right)^{\gamma-2}\left(\frac{E}{{\rm GeV}}\right)^{-\gamma}\exp\left(-\frac{E}{(1-E/E_{{\rm max}})E_{{\rm c}}}\right)\exp\left(-\frac{r^{2}}{r_{{\rm diff}}^{2}}\right),\label{eq:pg_solution}
\end{equation}
where $E_{{\rm max}}=(b_{0}t)^{-1}$ is the maximum energy of electron/positron
after propagation. The diffusion length $r_{{\rm diff}}$ is given
by $r_{\mathrm{diff}}(E,t)\approx2\sqrt{\lambda(E)D(E)t}$, where
$\lambda(E)=[1-(1-E/E_{\mathrm{max}})^{1-\delta}]/[(1-\delta)E/E_{\mathrm{max}}]$.

Electrons and positrons can also be produced via halo DM annihilation or decay.
In this work, we
shall focus on the DM annihilation as it is essential for all
the thermal relic models, and the extension of the analysis from DM annihilation
to DM decay is straightforward. DM particles in the Galactic halo may
annihilate into the standard model particles and make extra contributions
to the CR electron and positron fluxes. The source term of primary electrons and positrons from the annihilation
of Majorana DM particles takes the following form 
\begin{equation}
Q_{{\rm DM}}(\boldsymbol{r},p)=\frac{\rho(\boldsymbol{r})^{2}}{2m_{\chi}^{2}}\langle\sigma v\rangle\sum_{X}\eta_{X}\frac{dN^{(X)}}{dp},\label{eq:dm_source}
\end{equation}
where $\rho(\boldsymbol{r})$ is the DM energy density profile, $m_{\chi}$
stands for the DM particle mass, $\langle\sigma v\rangle$ is the velocity-weighted
annihilation cross section, $dN^{(X)}/dp$ is the injection energy
spectrum from DM particles annihilating into electrons and positrons via all possible
intermediate states $X$ and $\eta_{X}$ is the corresponding branching
fraction. The fluxes of CR electrons and positrons from DM annihilation depend
mildly on the choice of DM halo profile. In this work, we adopt the Einasto
profile \citep{Einasto:2009zd} 
\begin{equation}
\rho(\boldsymbol{r})=\rho_{\odot}\exp\left[-\left(\frac{2}{\alpha_{E}}\right)\left(\frac{r^{\alpha_{E}}-r_{\odot}^{\alpha_{E}}}{r_{s}^{\alpha_{E}}}\right)\right],
\end{equation}
with $\alpha_{E}\approx0.17$ and $r_{s}\approx20$ kpc. The local
DM energy density is taken to be $\rho_{\odot}=0.43\text{ GeV}\text{ cm}^{-3}$
\citep{Salucci:2010qr}. In this work, the injection spectra $dN^{(X)}/dp$
from DM annihilation are calculated using the numerical package \PYTHIA~\cite{Sjostrand:2007gs}.

It is known that DM annihilation as the dominant contribution to the
CR positron excess is severely constrained. The lack of excess in
the CR antiproton flux excludes the DM annihilation directly into
$q\overline{q}$, $W^{+}W^{-}$ and $Z^{0}Z^{0}$ final states. The
$e^{+}e^{-}$ channel leads to a very sharp spectral structure which
cannot fit the observed positron flux. The leptonic channel $\tau^{+}\tau^{-}$
channel is also ruled out by the Fermi-LAT data on the $\gamma$-rays from
the dwarf spheroidal galaxies (dSphs) \citep{Ackermann:2015zua}. Only
the $\mu^{+}\mu^{-}$ channel is marginally compatible with the positron data.
Note that the bump structure of the positron spectrum predicted from
the $\mu^{+}\mu^{-}$ annihilation is relatively narrow compared with
the broad excess observed by AMS-02. In this work, we shall only consider
DM as a subdominant component of the nearby sources, with the normalization
as a free parameter to be determined by the data.

As for the astrophysical explanations for the CR positron excess, 
two kinds of PWN explanations have been proposed and extensively discussed.
The one is considering a single nearby PWN as the major 
source of high energy $e^{\pm}$ \cite{Hooper:2017gtd,
Yuksel:2008rf,Pato:2010im,Hooper:2008kg,Yin:2013vaa,Wang:2017hsu}. 
The other one is considering the contributions from all known PWNe 
and usually assuming a common spectral index \cite{Malyshev:2009tw,
Grasso:2009ma,Delahaye:2010ji,DiBernardo:2010is,
DiMauro:2014iia,Manconi:2016byt,DiMauro:2017jpu,Manconi:2018azw}. 
In view of the possible TeV CRE break, we revisit
these two kinds of PWN hypotheses and their combination with 
a DM component with $\mu^{+}\mu^{-}$ final states.
\begin{itemize}
  \item \textbf{Model-A} (single PWN) We assume a generic nearby PWN with 
  the age $T_{\mathrm{psr}}$, distance $d_{\mathrm{psr}}$, spectral
  index $\gamma_{\mathrm{psr}}$, cut-off energy $E_{c,\mathrm{psr}}$,
  efficiency $\eta_{\mathrm{psr}}$ and spin-down luminosity $\dot{\mathcal{E}}_{\mathrm{psr}}$
  of the associated pulsar determined through fitting to data. Since the parameter $\eta_{\mathrm{psr}}$
  and $\dot{\mathcal{E}}_{\mathrm{psr}}$ are degenerate in the expression of $E_{\mathrm{tot,psr}}$,
  we take the product $\eta_{\mathrm{psr}}\dot{\mathcal{E}}_{\mathrm{psr}}$
  as a single parameter. Thus there are five free parameters \{$T_{\mathrm{psr}}$,
  $d_{\mathrm{psr}}$, $\gamma_{\mathrm{psr}}$, $E_{c,\mathrm{psr}}$,
  $\eta_{\mathrm{psr}}\dot{\mathcal{E}}_{\mathrm{psr}}$\} in this
  model.
  \item \textbf{Model-B} (Multiple PWNe) We consider the middle-aged PWNe
  with observed age $t_{{\rm obs}}$ in the range 50 kyr $<t_{{\rm obs}}<$
  $10^{4}$ kyr from the most updated ATNF catalog \citep{Manchester:2004bp}.
  We simply assume a common spectral index $\gamma$, efficiency $\eta$,
  and exponential cutoff $E_{c}$ for all PWNe. Thus there are three free 
  parameters \{$\gamma$, $\eta$, $E_{c}$\} in this model.
  \item \textbf{Model-C} (single PWN + DM) This model consists of all the
  primary and secondary astrophysical contributions as that in \textit{Model-A}
  plus the $e^{\pm}$ fluxes produced by DM with a typical $2\mu$ annihilation
  channel. The DM particle mass $m_{\chi}$ and cross section $\langle\sigma v\rangle$
  are allowed to vary freely in the global fit. Thus compared with 
  \textit{Model-A}, this model has two more free parameters.
  \item \textbf{Model-D} (multiple PWNe + DM) This model consists of all the
  primary and secondary astrophysical contributions as that in \textit{Model-B}
  plus the $e^{\pm}$ fluxes produced by DM with a typical $2\mu$ annihilation
  channel. The DM particle mass $m_{\chi}$ and cross section $\langle\sigma v\rangle$
  are allowed to vary freely in the global fit. Thus compared with 
  \textit{Model-B}, this model has two more free parameters.
\end{itemize}
For the primary and secondary CRE background components, three additional
free parameters \{$\gamma_{e}$, $\rho_{c}$,
$N_{e}$\} are introduced. We adopt the ``MED'' propagation
model and fit the parameters of the primary electron spectrum. 
We consider the CRE data from the five experiments (Fermi-LAT \cite{Abdollahi:2017nat}, 
AMS-02 \cite{Aguilar:2019ksn}, DAMPE \cite{Ambrosi:2017wek}, 
CALET \cite{Adriani:2018ktz}, and H.E.S.S. \cite{HESSICRC17}) and the 
CR positron data from AMS-02 \cite{Aguilar:2019owu}
with energy scale uncertainty included. Only the data with energies
above 25 GeV will be included in the fits to reduce the influence
of the solar modulation. In total 196 data points are included in
the analyses. We perform a Bayesian analysis 
to the data. To efficiently explore the high-dimensional parameter
space of the models, we adopt the MultiNest sampling algorithm 
\cite{Feroz:2007kg,Feroz:2008xx,Feroz:2013hea}. Details 
of the Bayesian statistical framework can be found
in \App{app:bayesian}. 


\begin{figure}[!t]
  \begin{center}
    \includegraphics[width=\columnwidth]{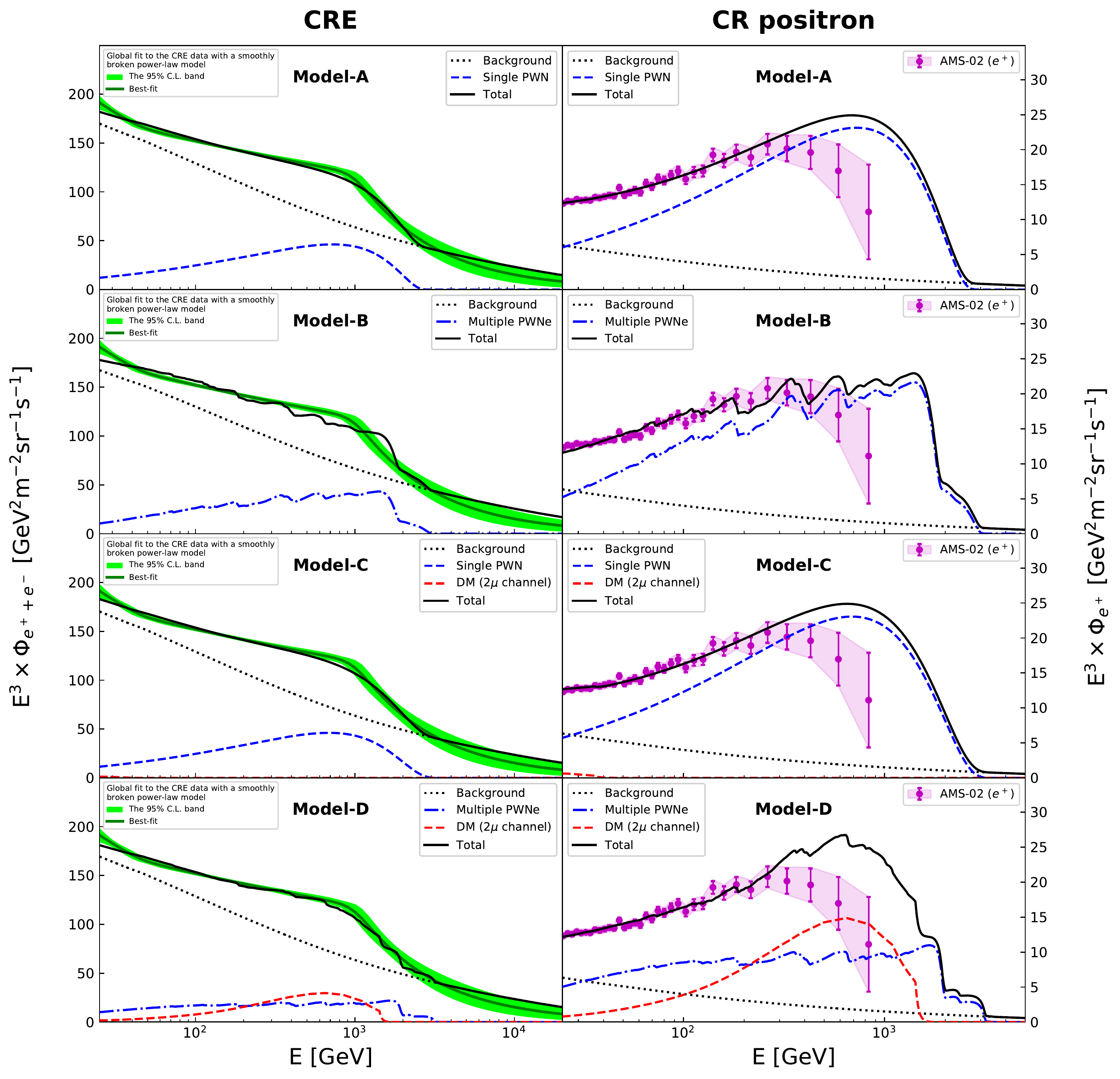}
  \end{center}
  \caption{CRE (left) and CR positron (right) fluxes for the best-fit 
  parameters of the charge-symmetric models (\textit{Model A, B, C, and D}) 
  described in \Sec{sec:c_sym_sources}. The black dotted, blue dashed, blue 
  dash-dotted, and red dashed curves represent the contribution from the 
  background, the single PWN, the middle-aged PWNe and the DM with a typical $2\mu$
  annihilation channel respectively. The black solid curve represents the sum 
  of all the components in each plot. The best-fit CRE flux and the 95\% C.L. 
  uncertainty band (green curve and band) from a global fit to the latest CRE 
  data with a smoothly broken power-law model are also illustrated for comparison.}
  \label{fig:Model_ABCD}
\end{figure}

The results of the fit are presented in \Fig{fig:Model_ABCD}.
As the figure shows, 
the TeV break measured in 
the CRE spectrum can be well reproduced by \textit{Model-A, -C and -D}, 
but difficult for \textit{Model-B}. 
For \textit{Model-C}, the fit result shows that the 
nearby PWN as a dominant source is favored. 
The contribution from DM annihilation is quite small. 
Compared with \textit{Model-A}, this model does not show
any improvement in the goodness-of-fit as the Bayes factor is 
around 1.2. Therefore, the additional DM component in this model is unnecessary.
For \textit{Model-D}, the fit result shows that the 
additional DM component can significantly improve the agreement with the data. 
Compared to \textit{Model-B}, the Bayes factor for \textit{Model-D} is greater than 150,
and the minimum $\chi^{2}$ value is reduced by 45.6 for two less degrees of freedom.

Although some models (\textit{Model-A and -D}) could give a successful fit with 
a $\chi^{2}/\mathrm{d.o.f.}\sim1.1-1.2$, we find that 
in general none of the models from A to D can well reproduce the CR
positron flux measured by AMS-02. This is related to the difference
in the energy regions where the CRE and CR positron spectra start to
soften. The steepening of the measured CRE spectra appears at around
TeV, while the dropoff of the measured positron spectrum appears at
around 300 GeV. The different behavior of the CRE and CR positron spectra makes
the models based on charge-symmetric $e^{\pm}$ sources in difficulty.

\subsection{Charge-asymmetric electron–positron sources \label{sec:c_asym_sources}}

In view of the difficulties in charge-symmetric source models, it seems that
to simultaneously account for both the CRE and CR positron flux data,
extra charge-asymmetric $e^{\pm}$ sources are needed.
SNR is a known source to contribute dominantly primary CR electrons only. 
Whereas,
the details of the release mechanism of electrons from SNRs are poorly known and 
still under debate \cite{Blasi:2013rva,Caprioli:2009fv,Ohira:2011xq,
Gabici:2009ak,Blasi:2011fi,Fornieri:2020xai}.
In this work, we adopt a simplified model, the burst-like injection model. 
The calculation of CR electrons from SNR is mostly the same as that of PWN.
The models based on a more realistic emission mechanism 
that electrons escape from SNR when its energy is larger than 
the maximal energy of electrons that can be confined in SNR 
have been extensively studied \cite{Ohira:2011xq,Blasi:2011fi,
Gabici:2009ak,Caprioli:2009fv}. 
In these models, the escape of electrons from SNR takes place 
in such a way that higher energy electrons escape earlier in the evolution, 
while lower energy ones leave later. 
The analysis in \cite{Gabici:2009ak} 
shows that the high energy electrons are emitted within a few thousand years 
from the supernova explosion. 
This timescale is much smaller than the age of sources typically considered 
to explain the CRE data at Earth, 
which supports the burst-like emission is a good approximation in our analysis.

In this section, we add a SNR component to the models previously considered.
We shall focus on the extension of \textit{Model-A} and \textit{Model-D}
as these two models fit the CRE data better than other models.
\begin{itemize}
  \item \textbf{Model-E} (single PWN + SNR) This model consists of all the primary
  and secondary astrophysical contributions as that in \textit{Model-A}
  plus the electron flux produced by a nearby SNR. The age $T_{\mathrm{snr}}$, 
  distance $d_{\mathrm{snr}}$, spectral index $\gamma_{\mathrm{snr}}$, cut-off 
  energy $E_{c,\mathrm{snr}}$, and the total energy emitted into 
  electrons $E_{\mathrm{tot,snr}}$ of the SNR are determined through 
  fitting to data. Thus compared with \textit{Model-A}, this model has five 
  more free parameters. 
  \item \textbf{Model-F} (multiple PWNe + DM + SNR) This model consists of all
  the astrophysical and DM contributions as that in \textit{Model-D}
  plus the electron flux produced by a nearby SNR. The age $T_{\mathrm{snr}}$, 
  distance $d_{\mathrm{snr}}$, spectral index $\gamma_{\mathrm{snr}}$, cut-off 
  energy $E_{c,\mathrm{snr}}$, and the total energy emitted into 
  electrons $E_{\mathrm{tot,snr}}$ of the SNR are determined through 
  fitting to data. Thus compared with \textit{Model-D}, this model has five 
  more free parameters.
\end{itemize}

\begin{figure}[t]
  \begin{center}
    \includegraphics[width=0.31\columnwidth]{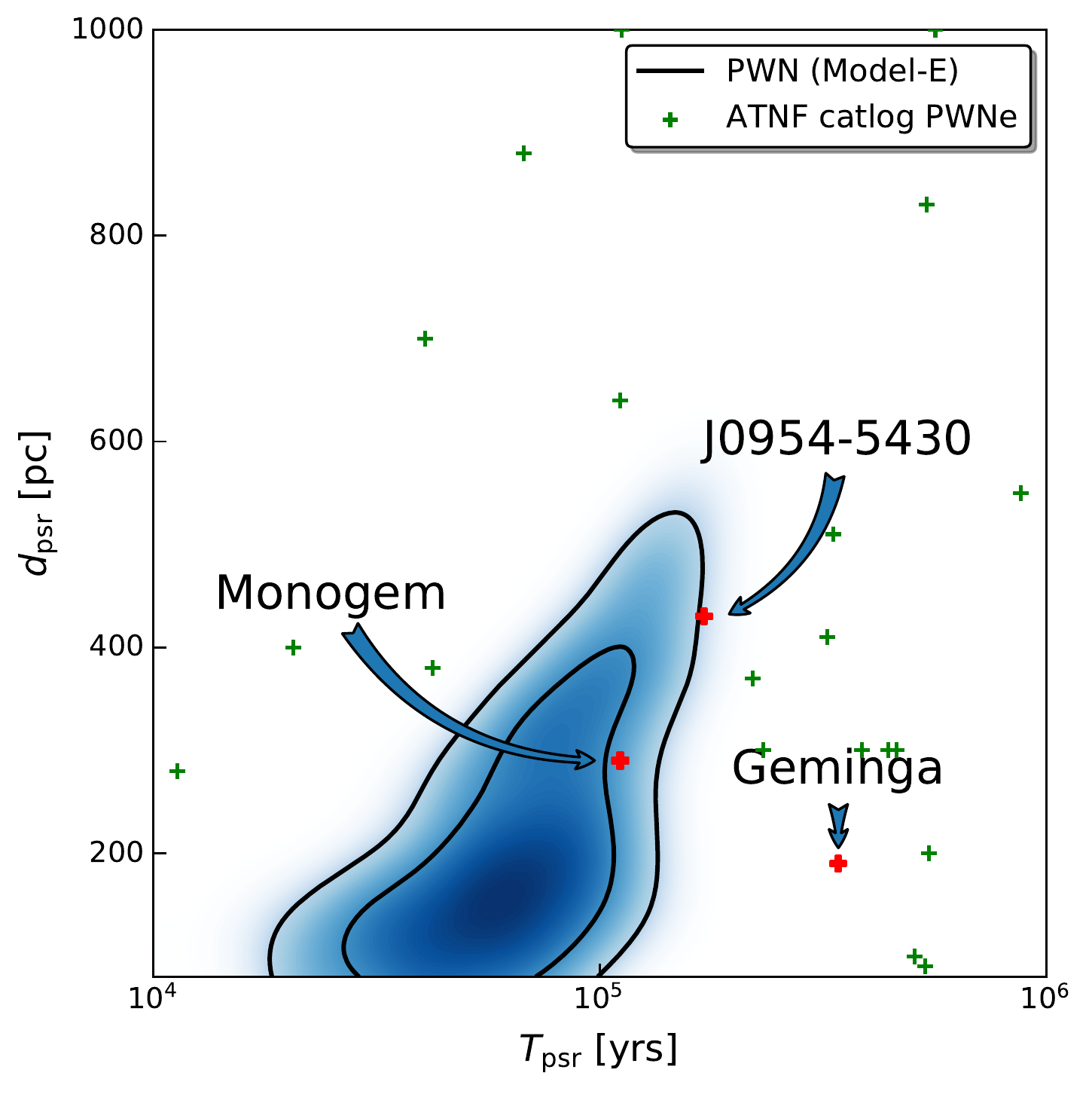}
    \includegraphics[width=0.30\columnwidth]{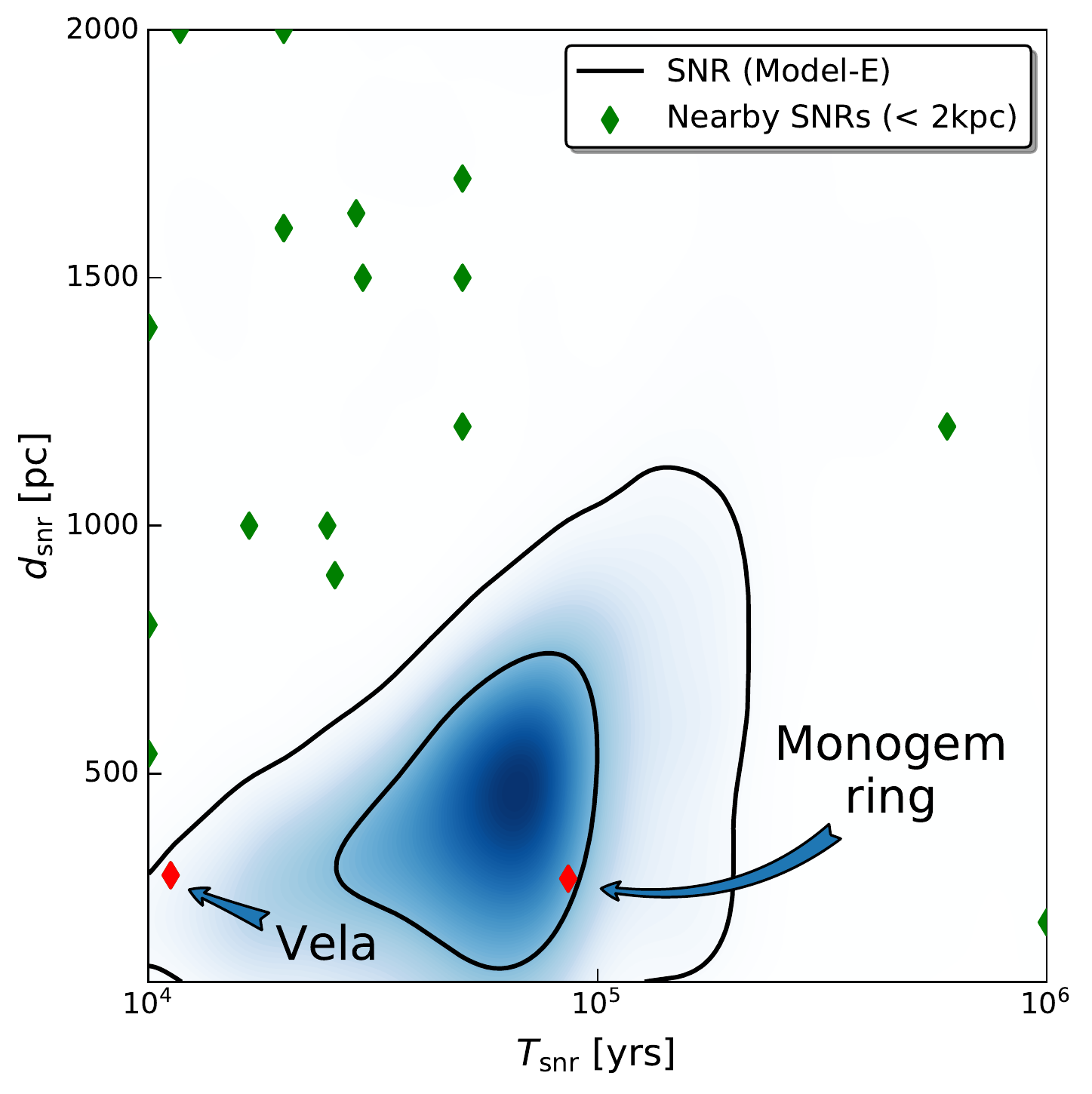}
    \includegraphics[width=0.37\columnwidth]{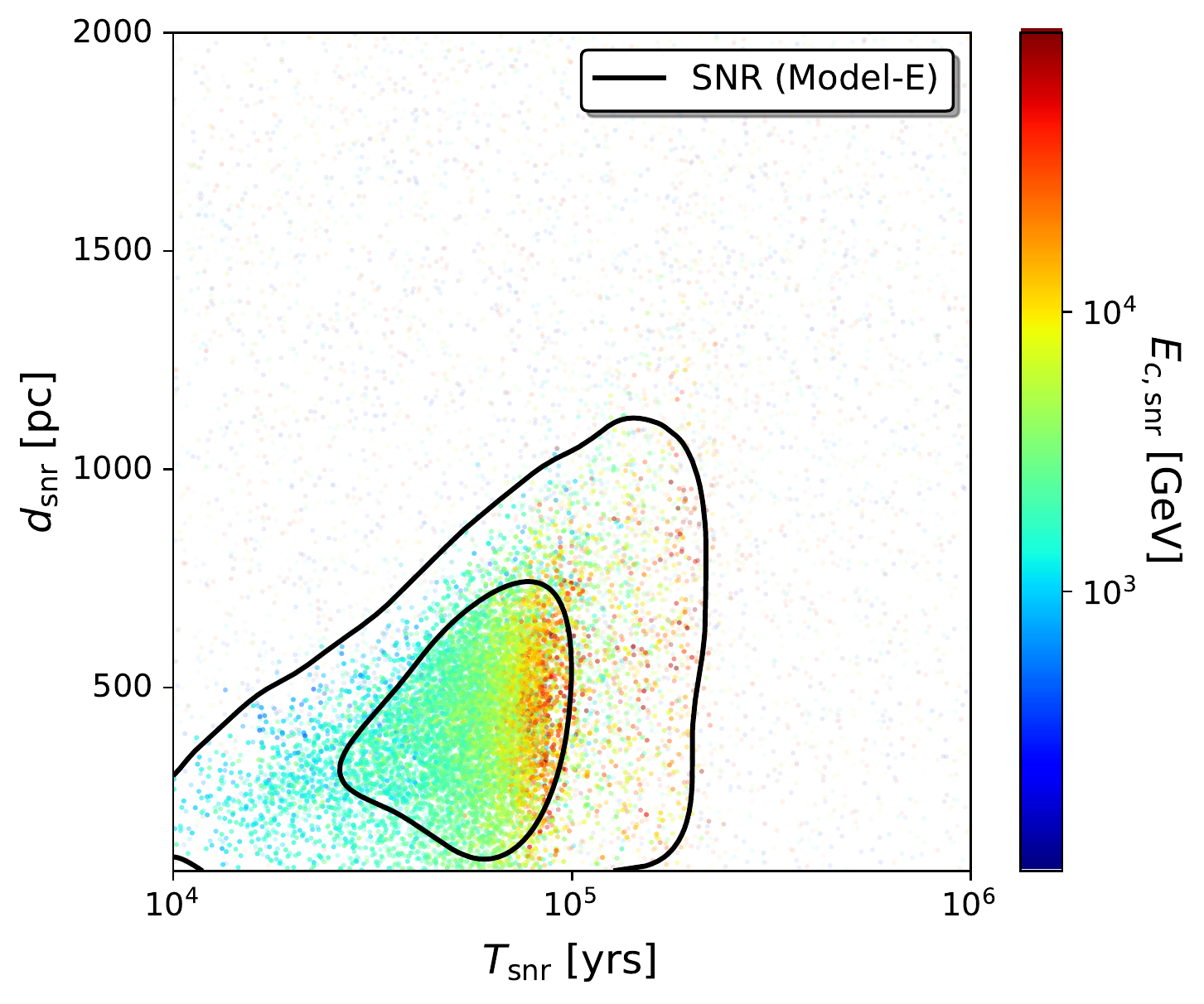}
  \end{center}
\caption{Results for \textit{Model-E} described in 
\Sec{sec:c_asym_sources}. Left panel shows the allowed regions at 68\% and 
95\% C.L. in ($T_{\mathrm{psr}}$, $d_{\mathrm{psr}}$) plane for the PWN in this
model, comparing with the PWNe listed in the ATNF catalog. Middle and Right 
panels show the allowed regions 68\% and 95\% C.L. in 
($T_{\mathrm{snr}}$, $d_{\mathrm{snr}}$) plane for the SNR in this model.
The nearby known SNRs (< 2 kpc) are illustrated in the middle panel for 
comparison. The scatter points in the right panel represent the posterior 
samples and are colored by the values of the cutoff energy of the SNR.}
\label{fig:singlePWNSNR_res}
\end{figure}

For \textit{Model-E}, the fit result shows that the 
additional SNR can significantly improve the agreement with the data. Compared 
with \textit{Model-A}, the Bayes factor for this model is 
$\sim 81$, and the minimum $\chi^{2}$ value is reduced by 23.9 for 
five less degrees of freedom. 
The additional SNR for this model turns out to be $\sim$ 57 kyr old,
located at $\sim$ 0.5 kpc from the Earth, with a spectral index $\sim$
2.2, an exponential cutoff energy $\sim$ 4.1 TeV, and a total energy
$\sim5.5\times10^{48}$ erg. Compared to the PWN in this model,
the additional SNR mainly contributes its electrons at around TeV.
In the middle panel of \Fig{fig:singlePWNSNR_res}, we plot
the allowed regions for the SNR in the ($T_{\mathrm{snr}}$, $d_{\mathrm{snr}}$)
plane at $68\%$ and $95\%$ C.L., together with the nearby known
SNRs ($<2~\mathrm{kpc}$) summarized in Table C.1. of \citep{Delahaye:2010ji}.
There are two SNRs, Vela and Monogem ring, falling in the regions allowed 
by the data. From the right panel of \Fig{fig:singlePWNSNR_res}, 
it can be seen that the SNRs falling in 
the allowed regions come in two kinds: one has an injection cutoff 
$E_{c,\mathrm{snr}}$ around several TeV, the other is sufficient old 
($T_{\mathrm{snr}}\ge10^{5}~{\rm {yrs}}$) which suffers from a cooling 
cutoff $E_{\mathrm{max,snr}}$ around TeV. Both ensure that the 
electron spectrum produced by the additional SNR drops sharply around TeV. 
The cutoff energy for the PWN in \textit{Model-E} is found to be $\sim0.8$
TeV, smaller than the one in \textit{Model-A} ($\sim3.2$ TeV), while
the spectral index $\gamma_{\mathrm{psr}}$ and the product 
$\eta_{\mathrm{psr}}\dot{\mathcal{E}}_{\mathrm{psr}}$
are similar to \textit{Model-A}. In \textit{Model-E}, less
electron-positron pairs are produced by the PWN in high energy region, 
which results in a good agreement to the CR positron spectrum. The corresponding
reduction in the CRE spectrum is offset by the electrons produced by the SNR.
In the left panel of \Fig{fig:singlePWNSNR_res}, we plot the allowed 
regions for the PWN in the ($T_{\mathrm{psr}}$, $d_{\mathrm{psr}}$) plane 
at $68\%$ and $95\%$ C.L. 
As the figure shows, the Monogem pulsar falls in the regions allowed by the data,
PSR J0954-5430 is on the edge of the regions and could be a possible candidate 
once the uncertainty on the determination of the distance and age of the pulsar 
is taken into account.

The Monogem pulsar is widely
considered to be a possible origin of the CR positron excess 
(see e.g.\citep{Hooper:2008kg,Hooper:2017gtd,Yuksel:2008rf,Yin:2013vaa}). 
However, the recent measurement of the surface brightness profile of TeV nebulae 
surrounding Monogem by HAWC \citep{Abeysekara:2017old}
suggests that the diffusion coefficient within a few tens of pc of 
these this pulsar is significantly lower than that expected in the ISM.
The HAWC collaboration claimed that such a low diffusion coefficient 
led to a negligible positron flux at Earth, disfavoring it as 
the source of the observed CR positron excess. 
In some two-zone diffusion models, the low diffusion region is only 
restricted to a small region close to the pulsar, and a larger
diffusion coefficient is possible outside the TeV nebula,
thus Monogem remains to be the best candidate
\citep{Hooper:2017tkg,Fang:2018qco,Profumo:2018fmz}. 
However, it was recently argued that after 
considering the GeV $\gamma$-ray observation of the nebula surrounding Monogem
provided by Fermi-LAT, Monogem was still disfavored
\citep{Shao-Qiang:2018zla,DiMauro:2019yvh}.

\begin{figure}[t]
  \begin{center}
    \includegraphics[width=0.43\columnwidth]{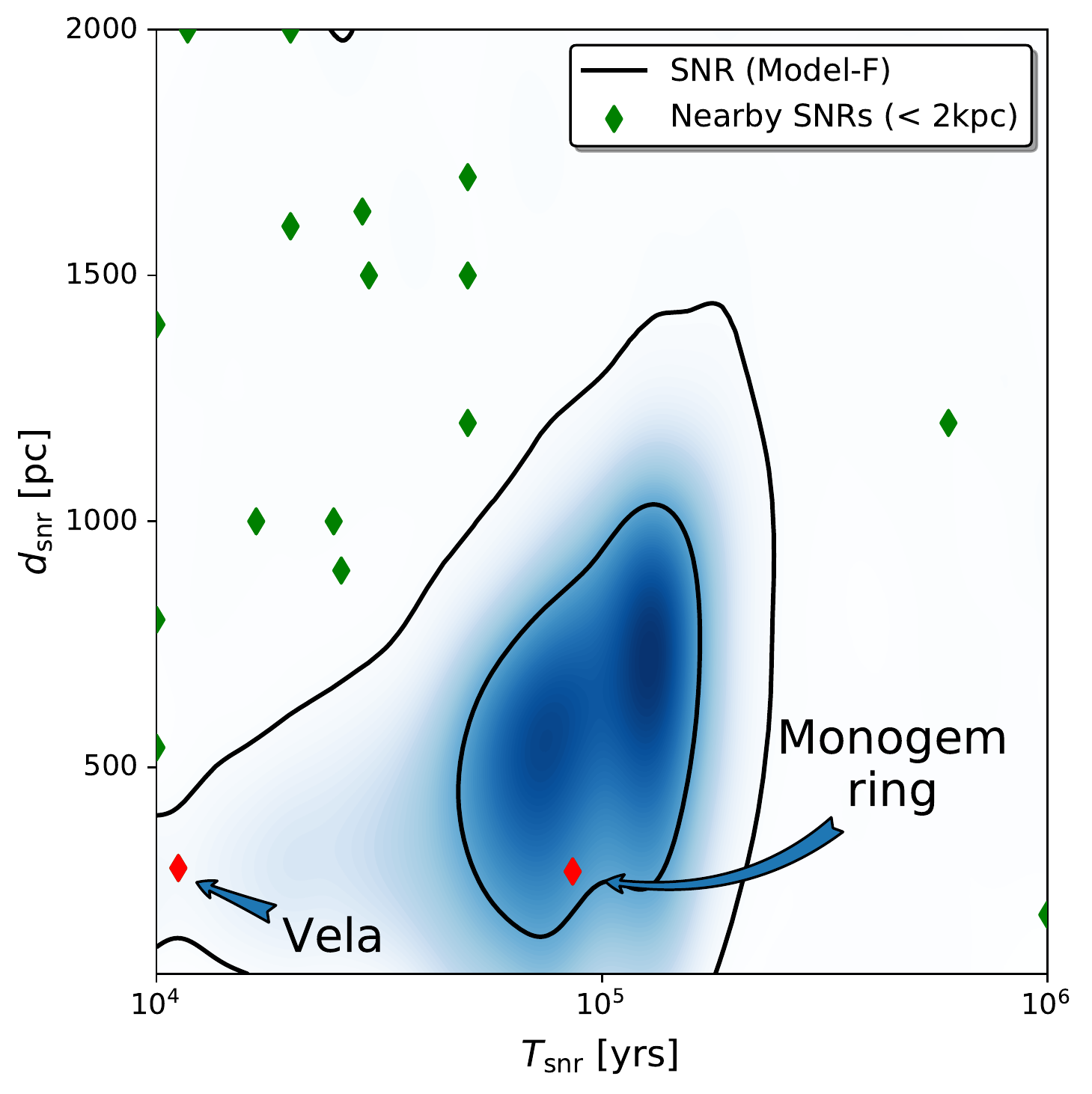}
  \end{center}
  \caption{The allowed regions at 68\% and 95\% C.L. in 
  ($T_{\mathrm{psr}}$, $d_{\mathrm{psr}}$) plane for the SNR in 
  \textit{Model-F}, comparing with the known SNRs within 2 kpc.}
  \label{fig:catalogPWNDMSNR_res2}
\end{figure}

\begin{figure}[t]
  \begin{center}
    \includegraphics[width=0.43\columnwidth]{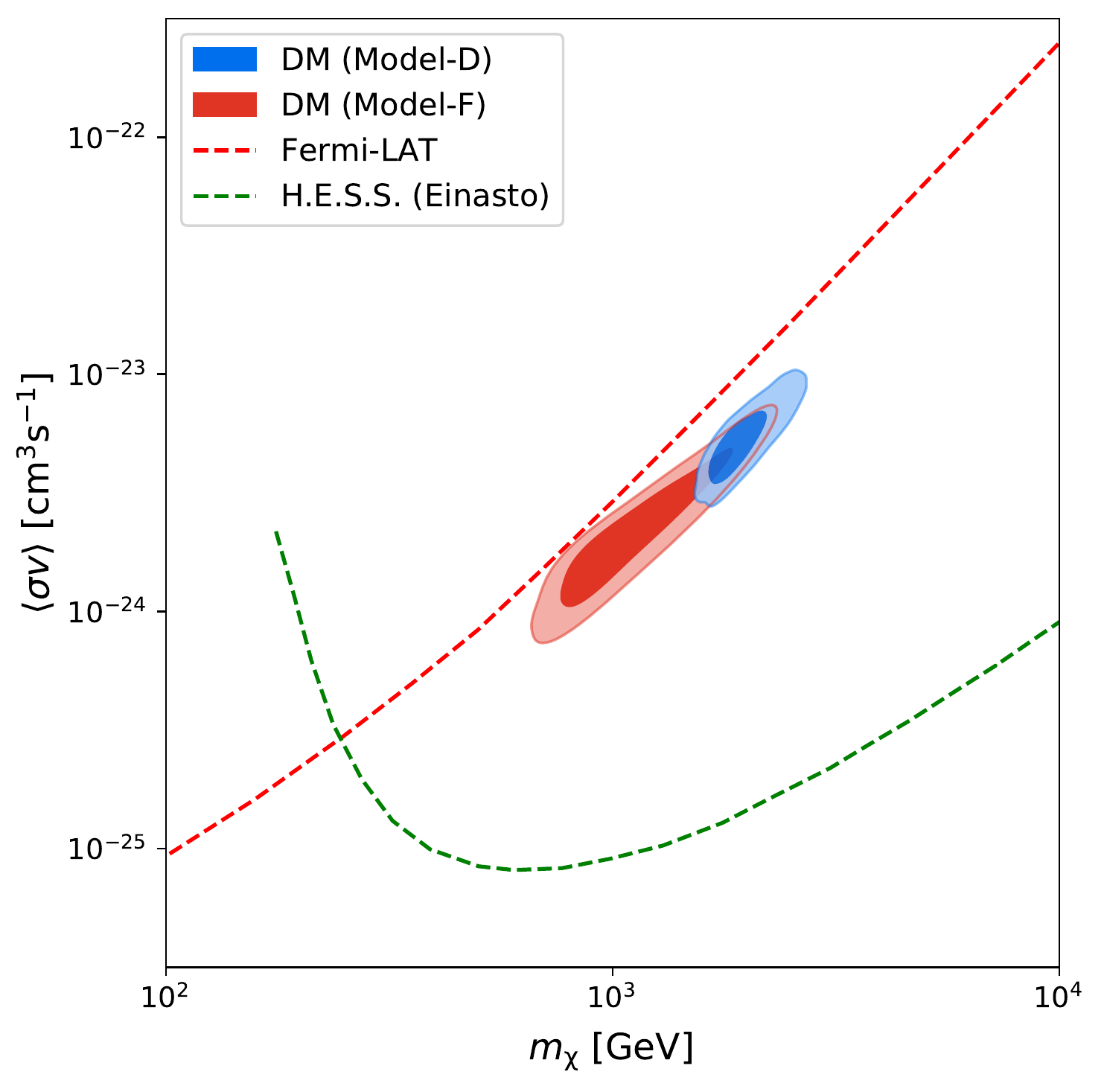}
  \end{center}
  \caption{The allowed regions at $68\%$ and $95\%$ C.L. in 
  ($m_\chi$, $\langle \sigma v \rangle$) plane for the DM in 
  \textit{Model-D} (blue contours) and in \textit{Model-F} (red contours). 
  The $95\%$ C.L. upper-limits given by Fermi-LAT \cite{Ackermann:2015zua} 
  (red dashed curve) and H.E.S.S. \cite{Abdallah:2016ygi} (green dashed curve) 
  are also plotted for comparison.}
  \label{fig:catalogPWNDMSNR_res1}
\end{figure}

For \textit{Model-F}, the additional SNR can also
significantly improve the agreement with the data. Compared with 
\textit{Model-D}, the Bayes factor for this model is 
$\sim 22$, and the minimum $\chi^{2}$ value is reduced by 25.4 for 
five less degrees of freedom. 
The additional SNR for this model turns out to be $\sim$ 152 kyr old,
located at $\sim$ 0.54 kpc from the Earth, with a spectral index
$\sim$ 1.9, and a total energy $\sim4.1\times10^{48}$ erg. The maximum 
energy of electrons surviving from cooling process for the SNR is $\sim 1.47$ 
TeV. Compared to the multiple PWNe and the DM component in this model, 
the additional SNR mainly contributes its electrons at around TeV. 
In \Fig{fig:catalogPWNDMSNR_res2}, we plot the allowed regions for 
the SNR in the ($T_{\mathrm{snr}}$, $d_{\mathrm{snr}}$) plane at 
$68\%$ and $95\%$ C.L, together with the nearby known SNRs ($<2~\mathrm{kpc}$) 
summarized in Table C.1. of \citep{Delahaye:2010ji}. There are two SNRs, 
Vela and Monogem ring, falling in the regions allowed by the data. The DM 
particle in this model turns out to be with mass $\sim$ 1 TeV and 
cross section $\sim1.69\times10^{-24}~{\rm cm^{3}s^{-1}}$, which is less massive 
than that in \textit{Model-D} ($\sim$ 1.9 TeV), while the favored 
parameters for the middle-aged PWNe are similar to \textit{Model-D}.
In \textit{Model-F}, less electron-positron pairs are produced by the DM in 
high energy region, which resulting in a good agreement to the CR positron 
spectrum. The corresponding reduction in the CRE spectrum is offset by the 
electrons produced by the additional SNR. 
In \Fig{fig:catalogPWNDMSNR_res1}, we plot the allowed regions 
at 68\% and 95\% C.L. in ($T_{\mathrm{psr}}$, $d_{\mathrm{psr}}$) plane for 
the DM of both \textit{Model-D and -F}.
As the figure shows, the allowed regions are consistent with the 
limits derived from Fermi-LAT data of $\gamma$-rays from dwarf 
galaxies, but in tension with the H.E.S.S. data from the GC. 
It is necessary to note that the constraints derived from the GC 
observation suffer from large uncertainty of the DM density around 
the GC. For instance, for a cored DM density with 
a core radius of 500 pc, the limits are one order of magnitude 
weaker \citep{HESS:2015cda}, while changing the DM density around
the GC will not significantly affect the fit to the CRE and CR positron data.
The constraints derived from the dwarf galaxies are not sensitive to 
the choice of the DM profile. 

For both \textit{Model-E and -F}, the fit results show that a large total 
injection energy $\log_{10}(E_{\rm tot}/{\rm erg})=48.0 \pm 0.7$ of 
the additional SNR is favored by data. 
For a standard Supernova (SN) explosion event 
that carries $O(10^{51})$ erg of kinetic energy, 
this corresponds to a conversion efficiency into electrons 
about $\log_{10}f = -3.0 \pm 0.7$, which is consistent with the limits given
in the literature \cite{Delahaye:2010ji} within uncertainties.

We have also considered other possibilities such as 
the combination of multiple PWNe plus a SNR.
We find that the inclusion of the additional SNR can significantly 
improve the agreement with the CRE data, 
but have no visible effect on the prediction of CR positron flux. 
The predicted positron spectrum in this model is similar to that of 
\textit{Model-B (multiple PWNe)}, thus can not well explain the data either.
The additional SNR for this model turns out to be favored with age 
$\sim$ 12 kyr old, located at $\sim$ 0.19 kpc from the Earth, a spectral index
$\sim$ 1.7, a cutoff energy $\sim 492$ GeV, and a total injection 
energy $\sim1.1\times10^{47}$ erg. 
Compared to the multiple PWNe components in this model, 
the additional SNR mainly contributes its electrons at around several hundred GeV.
Since this model can not well reproduce the CR positron flux measured 
by AMS-02, We will not discuss it further.

\begin{figure}[t]
  \begin{center}
    \includegraphics[width=\columnwidth]{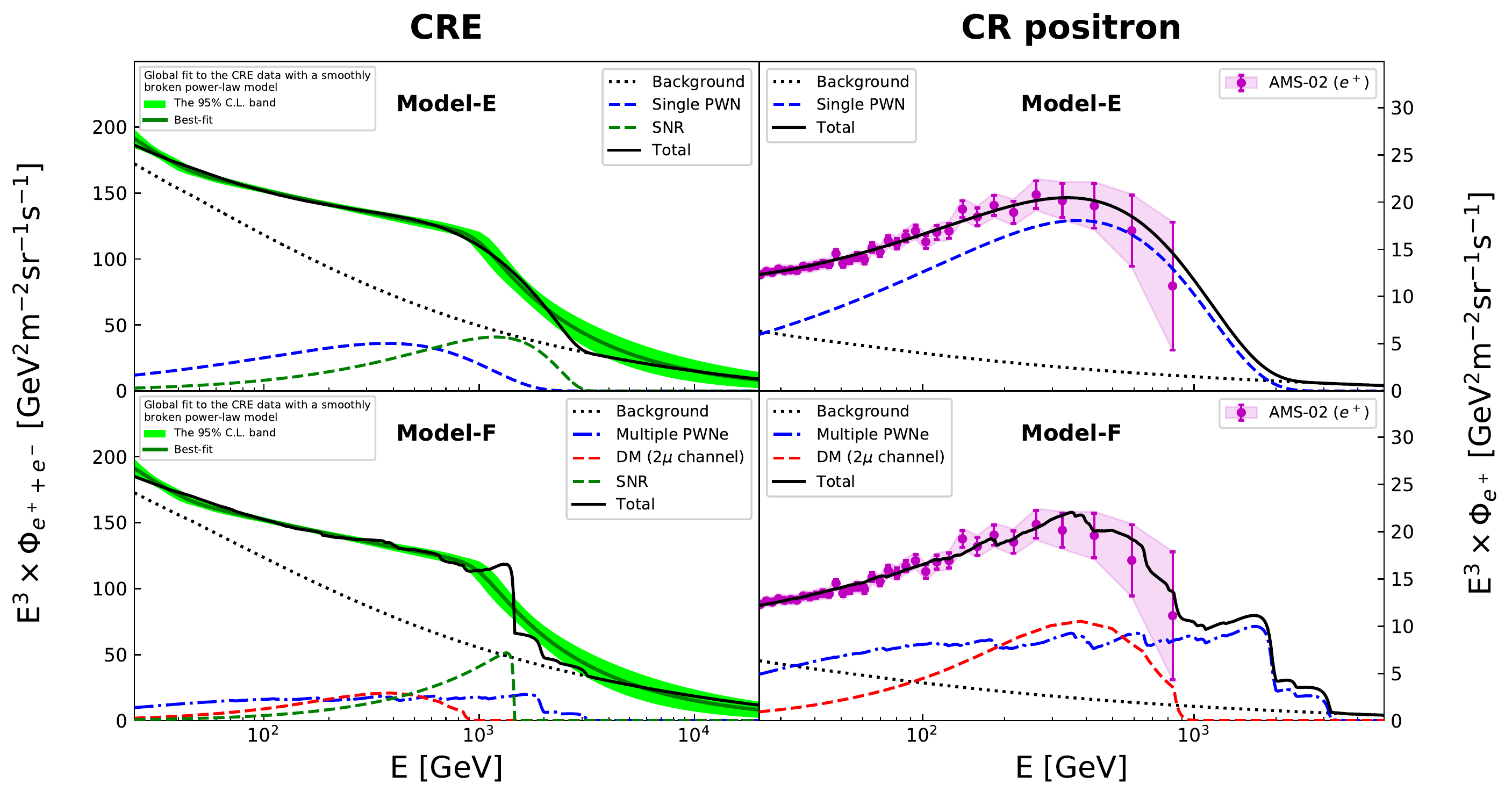}
  \end{center}
  \caption{Same as \Fig{fig:Model_ABCD}, but for the charge-asymmetric models 
  (\textit{Model-E and -F}) described in \Sec{sec:c_asym_sources}. The contribution 
  from the additional SNR is represented by the green-dashed curve.}
  \label{fig:Model_EF}
\end{figure}



\begin{figure}[t]
  \begin{center}
    \includegraphics[width=0.47\columnwidth]{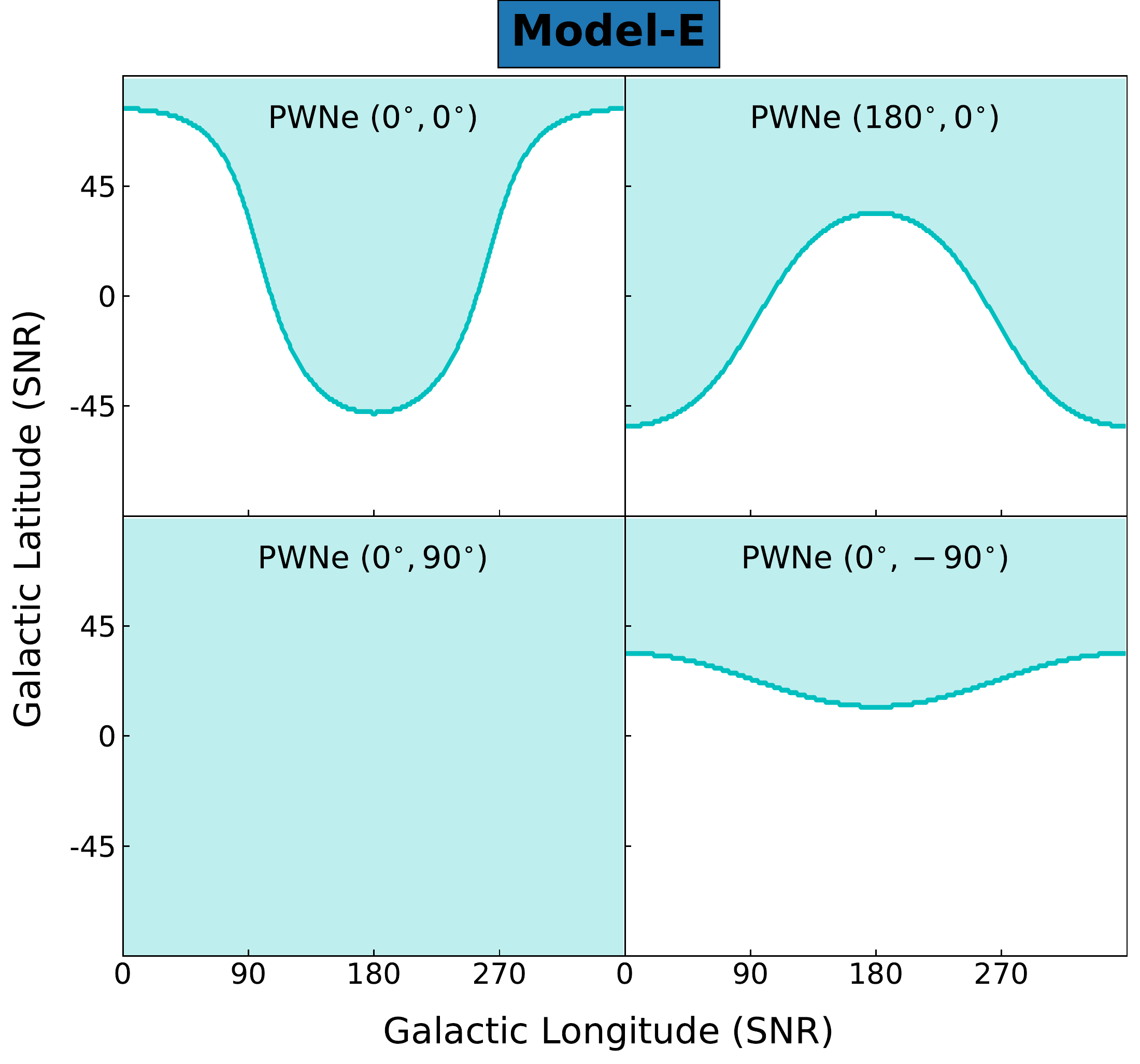}
    \includegraphics[width=0.48\columnwidth]{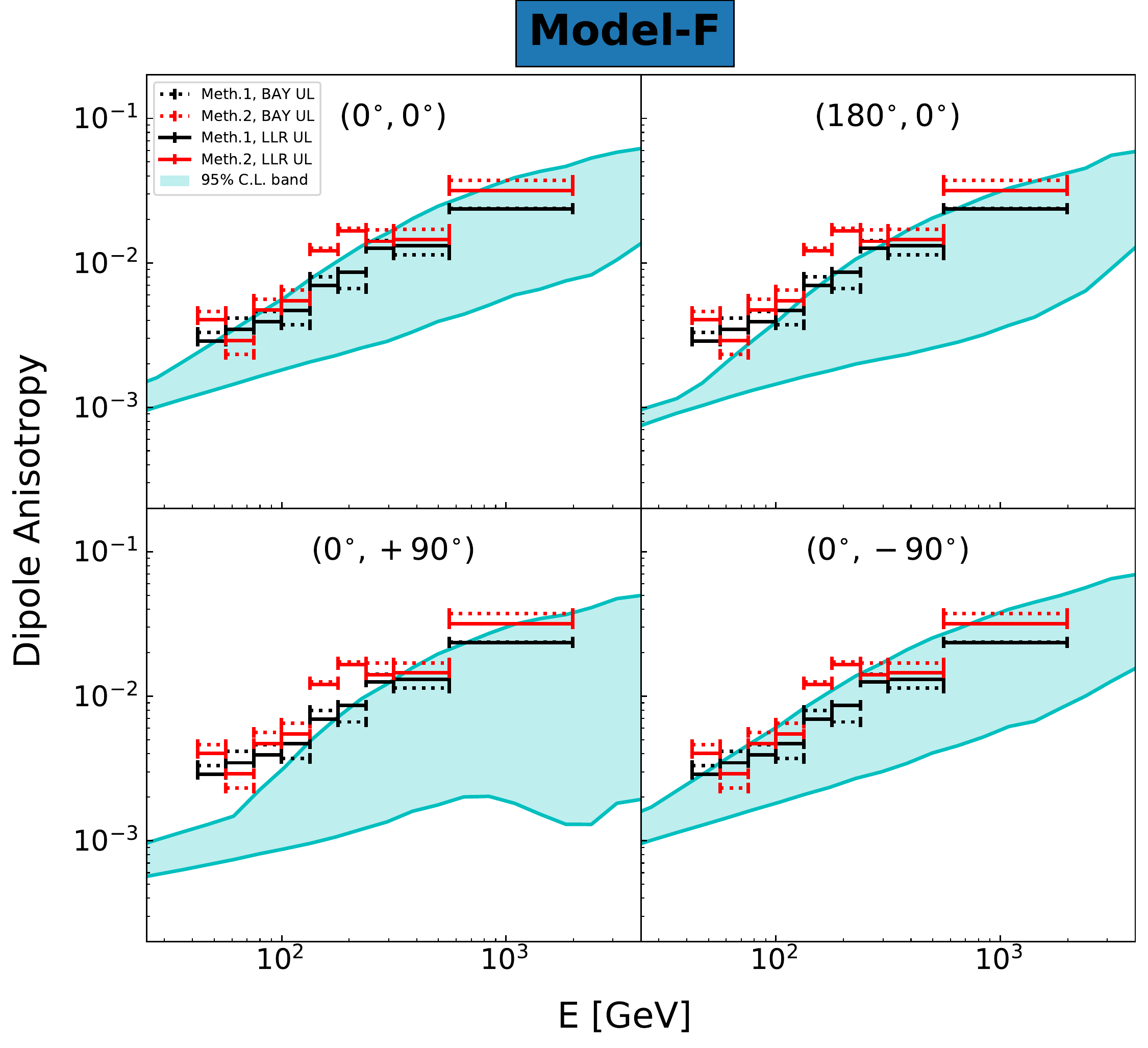}
  \end{center}
  \caption{Left panel illustrates the allowed regions for the position of the 
  SNR in \textit{Model-E}, which are derived by comparing the predicted CRE 
  dipole anisotropy from the best fit to the CRE and the CR positron 
  data with the upper-limits given by Fermi-LAT \cite{Abdollahi:2017kyf}.
  The PWN in this model is assumed to be in the position with Galactic coordinate,
  $\left(0^{\circ}, 0^{\circ}\right)$, $\left(180^{\circ}, 0^{\circ}\right)$, 
  $\left(0^{\circ}, +90^{\circ}\right)$ and $\left(0^{\circ}, -90^{\circ}\right)$
  respectively.
  Right panel illustrates the predictions for the dipole anisotropy of the total 
  CRE flux within the 95\% C.L. uncertainty band derived from fitting to the 
  CRE and the CR positron data for \textit{Model-F}. The SNR in this model is 
  assumed to be in the position with Galactic coordinate, 
  $\left(0^{\circ}, 0^{\circ}\right)$, $\left(180^{\circ}, 0^{\circ}\right)$, 
  $\left(0^{\circ}, +90^{\circ}\right)$ and $\left(0^{\circ}, -90^{\circ}\right)$ 
  respectively. The 95\% C.L. upper-limits given by Fermi-LAT 
  \cite{Abdollahi:2017kyf} are also plotted for comparison.}
  \label{fig:anisotropy_ME_MF_res}
\end{figure}

In addition, we compute the dipole anisotropy of CRE flux predicted by 
\textit{Model-E and -F} by using the method described in \Eq{aniso_sum} and 
then compare the results with the existing upper limits from Fermi-LAT
\cite{Abdollahi:2017kyf}. 
The anisotropies from the CRE background component and the DM component are 
computed by \Eq{aniso2} with the GALPROP code. The anisotropy from discrete 
PWN/SNR sources is given
\begin{equation}
  \Delta_{s}(E)=\frac{3}{2c \lambda(E)} \frac{|\boldsymbol{r}|}{t}, \label{eq:anisotropy_SNR}
\end{equation}
which is obtained from \Eq{eq:pg_solution}.
The term $\lambda(E)$ is insensitive to the value of $\delta$.
Varying the value of $\delta$ from 0.3 to 0.5 
for a typical $10^5$ year PWN/SNR, 
the uncertainty in $\lambda(E)$ is found to be within 5\%. 

For \textit{Model-E}, it is no simple task to give the 
total CRE anisotropy as the positions of the PWN 
and the SNR in this model are arbitrary. 
To have a clear view of how this 
model is constrained by the current CRE anisotropy observations, we 
pick up four representative positions with Galactic coordinate, 
$\left(0^{\circ}, 0^{\circ}\right)$, $\left(180^{\circ}, 0^{\circ}\right)$, 
$\left(0^{\circ}, +90^{\circ}\right)$ and $\left(0^{\circ}, -90^{\circ}\right)$ 
for the PWN and scan the position of the SNR. We compute the predicted 
CRE anisotropy (\textit{averaged} over the energy bins given by Fermi-LAT
\cite{Abdollahi:2017kyf}) for the best-fit parameters listed in 
\Tab{tab:singlePWNSNR}. We compare our predictions to the Fermi-LAT
upper-limits on the CRE anisotropy, choosing the more constraining 
results named Bayesian Method 1 in \cite{Abdollahi:2017kyf}.
Whenever our predictions overestimate one data point, we consider the position
pair for the PWN and the SNR as excluded. In the left panel of 
\Fig{fig:anisotropy_ME_MF_res}, we illustrate the position pairs surviving 
from the Fermi-LAT limits. As the figure shows, the CRE anisotropy observations 
could indeed provide a valuable information on the positions 
of the PWN and SNR. 

For \textit{Model-F}, only the position of the SNR 
is arbitrary. Here, we pick up four representative positions with Galactic 
coordinate, 
$\left(0^{\circ}, 0^{\circ}\right)$, $\left(180^{\circ}, 0^{\circ}\right)$, 
$\left(0^{\circ}, +90^{\circ}\right)$ and $\left(0^{\circ}, -90^{\circ}\right)$ 
for the SNR and compute the CRE anisotropy 
for all the configurations within 95\% C.L. selected by fitting to the CRE and CR
positron spectra. In the right panel of \Fig{fig:anisotropy_ME_MF_res}, 
we illustrate our anisotropy predictions,
together with the upper-limits given by Fermi-LAT. As the figure shows, 
our predictions reach the limits given by Fermi-LAT. 
Thus, the dipole anisotropy in the CRE arrival direction could set
additional constraints to this model.

The predicted CRE anisotropies of both \textit{Model-E and -F} reach 
the current Fermi-LAT limits and some parameter spaces are excluded.
The future Chinese Space Station based instrument HERD 
is expected to have a better capability of anisotropy detection than Fermi-LAT 
due to a better energy resolution for electrons and 
a better electron/proton separation power \cite{Zhang:2014qga}. 
It is planned for operation starting around 2025 for about 10 years
and may provide insights into these two models.

\section{discussions/conclusions}\label{sec:discuss}
In this work, we have employed the ``MED'' model as a benchmark model, 
in which the power-law index in the diffuse coefficient is $\delta \approx 0.3$.
The value of $\delta$ in the ``MED'' model is very close to 1/3 from the 
Kolmogorov diffusion \cite{1991RSPSA.434....9K}. 
Moreover, the latest AMS-02 B/C data at high rigidities 
is well described by a single power with 
index $\Delta=0.333 \pm 0.014(\mathrm{fit}) \pm 0.005(\mathrm{syst})$, 
which is in good agreement with the Kolmogorov diffusion. 
The value of $\delta$ depends on the analysis framework, e.g.
the recent analysis based on a semi-analytical approach 
yields $\delta \approx 0.43 - 0.53$ \cite{Genolini:2019ewc},
the analyses based on the numerical code DRAGON and GALPROP 
yield a value of 0.45 \cite{Fornieri:2019ddi} and 0.36 \cite{Xue:2019esp}, 
respectively. 


For an estimation of the uncertainties in CR propagation model, 
we consider three representative propagation models, 
the ``MIN'', ``MED'', and ``MAX'' models in \cite{Jin:2014ica}.
These models correspond to typical change due to the height of 
the propagation halo.
We have checked that varying the main propagation parameters 
in the allowed ranges found in \cite{Jin:2014ica}
the electron/positron fluxes are included between 
the fluxes obtained for the ``MIN'' and ``MAX'' propagation models.
The choice of different propagation models can result in the change of the 
secondary electron/positron flux up to 25\%, 
but has no significant effect on the primary electron flux.
For the total CRE background components, 
including both the primary and secondary electrons/positrons, 
the change of the flux is only within a few percent levels. 
The change in the propagation models only leads to minor changes
in our analysis, since the secondary electrons and positrons are subdominant
in the high energy region. 
%
For the DM-induced electrons/positrons, we checked 
the ``MIN'', ``MED'', and ``MAX'' models and 
found that the uncertainty in the positron/electron flux from 
DM annihilation into $2\mu$ is within a factor of 1.5.
Rescaling the favored DM parameter spaces found in Model-F 
with the ``MED'' model (see in \Fig{fig:catalogPWNDMSNR_res1}) 
by multiplying $\langle \sigma v \rangle$ by a factor of 1.5,
most parameter spaces are still under the upper-limits given by Fermi-LAT.
In this work, the propagation of the electrons/positrons from 
discrete PWN/SNR is calculated by an analytical approach 
with a spherically symmetric infinite boundary condition,
and the fluxes of that are related to the propagation parameters 
$D_0$ and $\delta$ through the diffusion length $r_{\rm{diff}}$ 
(see in \Eq{eq:pg_solution}).
Changing the propagation model from ``MED'' to ``MIN'' (``MAX'') model
can lead to a rescaling of $r_{\rm{diff}}$ by a factor of 0.74 (1.28).
To reproduce the CRE and CR positron data, the favored distance $d_{\rm psr(snr)}$
and total energy that emitted into electrons and positrons $E_{\rm tot,psr(snr)}$ 
of the PWN/SNR are expected to be rescaled by a factor of 0.74 (1.28) 
and 0.41 (2.1), respectively.

The CRE anisotropy in the arrival direction mainly comes from nearby sources 
(PWN/SNR), which mainly depends on the age and distance of the source
(see in \Eq{eq:anisotropy_SNR}).
As mentioned above, the change in the propagation models can lead to 
the change of source distance by about 27\%, 
thus a similar amount of change is expected in the CRE anisotropy.

In this paper, we have performed a global analysis to the latest CRE data
including Fermi-LAT, AMS-02, CALET, DAMPE, and H.E.S.S. We showed that 
a consistent fit of all the five data sets can be achieved by including 
the absolute energy scale uncertainties of each experiment. The global 
fit result strongly favors the existence of a break at $\sim 1$ TeV. 
After the break, the CRE spectral power index softens from 
$\sim 3.10$ to $\sim 3.89$, which confirms the result of DAMPE at a higher 
significance $\sim 13.3~\sigma$.

In view of the tentative CRE break, we have revisited a number
of models of nearby sources, such as a single generic PWN, known
multiple PWNe from the ATNF catalog, and their combinations with 
either an additional DM component or a SNR. We showed that the 
CRE break at $\sim 1$ TeV, together with the CR positron spectrum 
peaking at $\sim 300$ GeV points towards the possibility that the 
nearby sources are highly charge asymmetric. Among the models under 
consideration, only the model with a PWN plus SNR 
(labeled \textit{Model-E} in our paper) and the model with all 
middle-aged PWNe plus a SNR and a DM component which annihilates 
directly into $2\mu$ (labeled \textit{Model-F} in our paper), can well 
account for the current CRE and CR positron spectra simultaneously. 
For \textit{Model-E},
the data favor a nearby middle-aged PWN with a spectral index $\sim2$
and an energy cutoff at $\sim0.8$ TeV. Possible PWN candidates include
Monogem and PSR J0954-5430, while Monogem is excluded by the observation
of HAWC. The favored additional SNR turns out
to have a spectral index $\sim2.2$ and a total energy $\sim5.5\times10^{48}$
erg. Possible SNR candidates include Vela and Monogem ring. For \textit{Model-F},
the data favor a DM particle with mass $\sim$ 1 TeV and annihilation
cross-section $\sim1.69\times10^{-24}~{\rm cm^{3}s^{-1}}$. The favored
parameters are consistent with the limits derived from Fermi-LAT data
of $\gamma$-rays from dwarf galaxies,
but in tension with the H.E.S.S. data from the GC. 
The middle-aged PWNe in this model turn out to have a spectral $\sim2$, 
and an efficiency $\sim0.098$.
The favored additional SNR for this model turns out to have a
spectral index $\sim1.9$ and a total energy $\sim4.1\times10^{48}$
erg. Possible SNR candidates include Vela and Monogem ring.
In addition, we calculated the predicted dipole anisotropy on
CRE flux for both models and compared it with the present upper-limits
given by Fermi-LAT. We showed that the present Fermi-LAT data on 
the CRE anisotropy could be useful in understanding the
properties of the $e^{\pm}$ sources in our models.

In the near future, with increased statistics and improved understanding of the 
detector's performance, more consistent measurements of CRE flux among different
experiments might be achieved, which will provide remarkable insights into the 
models tested in the present analysis. 

\section{Acknowledgment}
This work is supported in part by the National Key R\&D Program of China 
No.~2017YFA0402204 and by the National Natural Science Foundation of China (NSFC) 
No.~11825506, 
No.~11821505, 
No.~U1738209, 
No.~11851303 
and No.~11947302.

\bibliographystyle{arxivref}
\bibliography{CRE_break}

\appendix
\section{Systematic uncertainty on the flux due to the energy scale uncertainty}\label{app:flux_correction}
\begin{figure}[t]
  \centering
  \begin{center}
    \includegraphics[width=0.46\columnwidth]{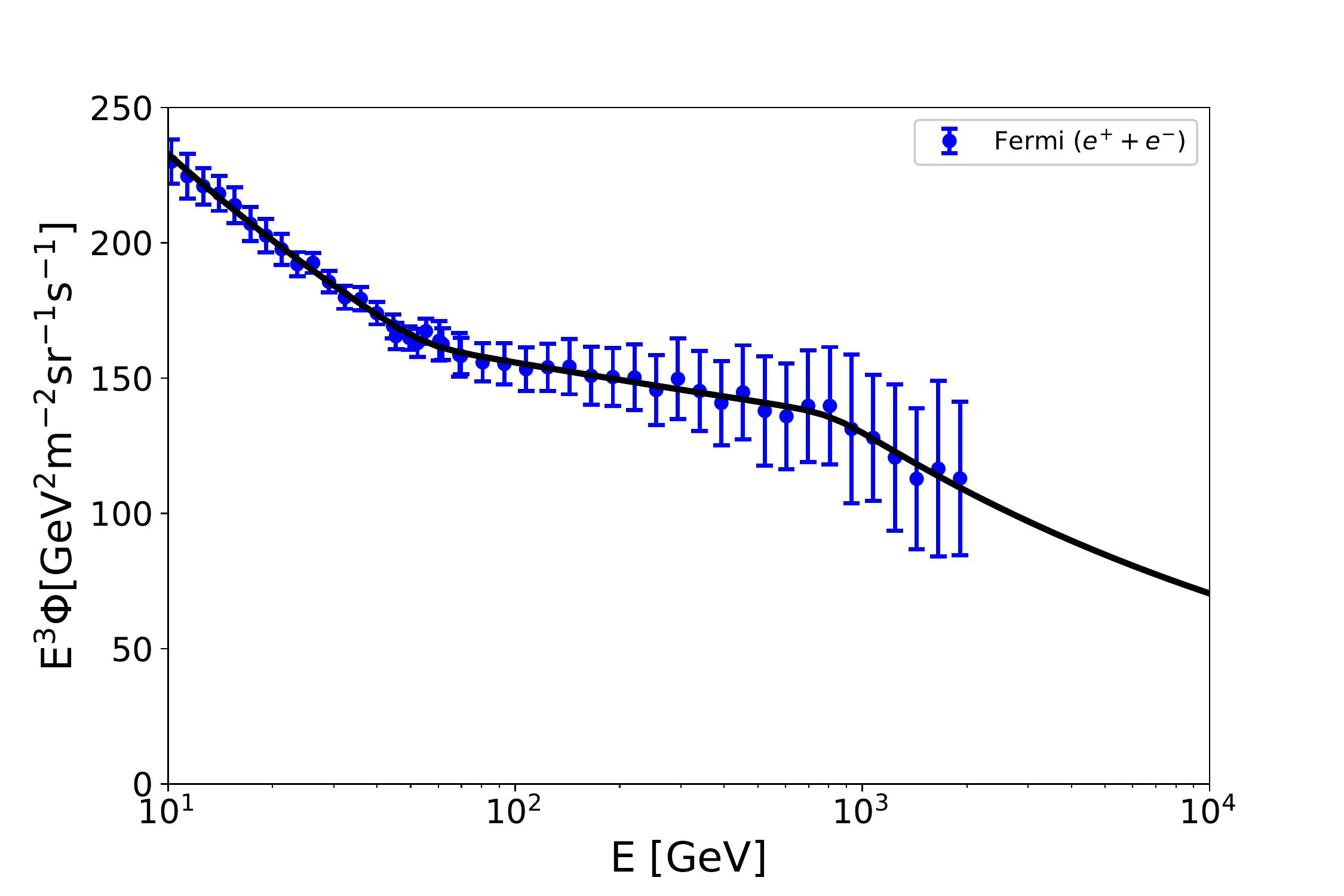}
    \includegraphics[width=0.46\columnwidth]{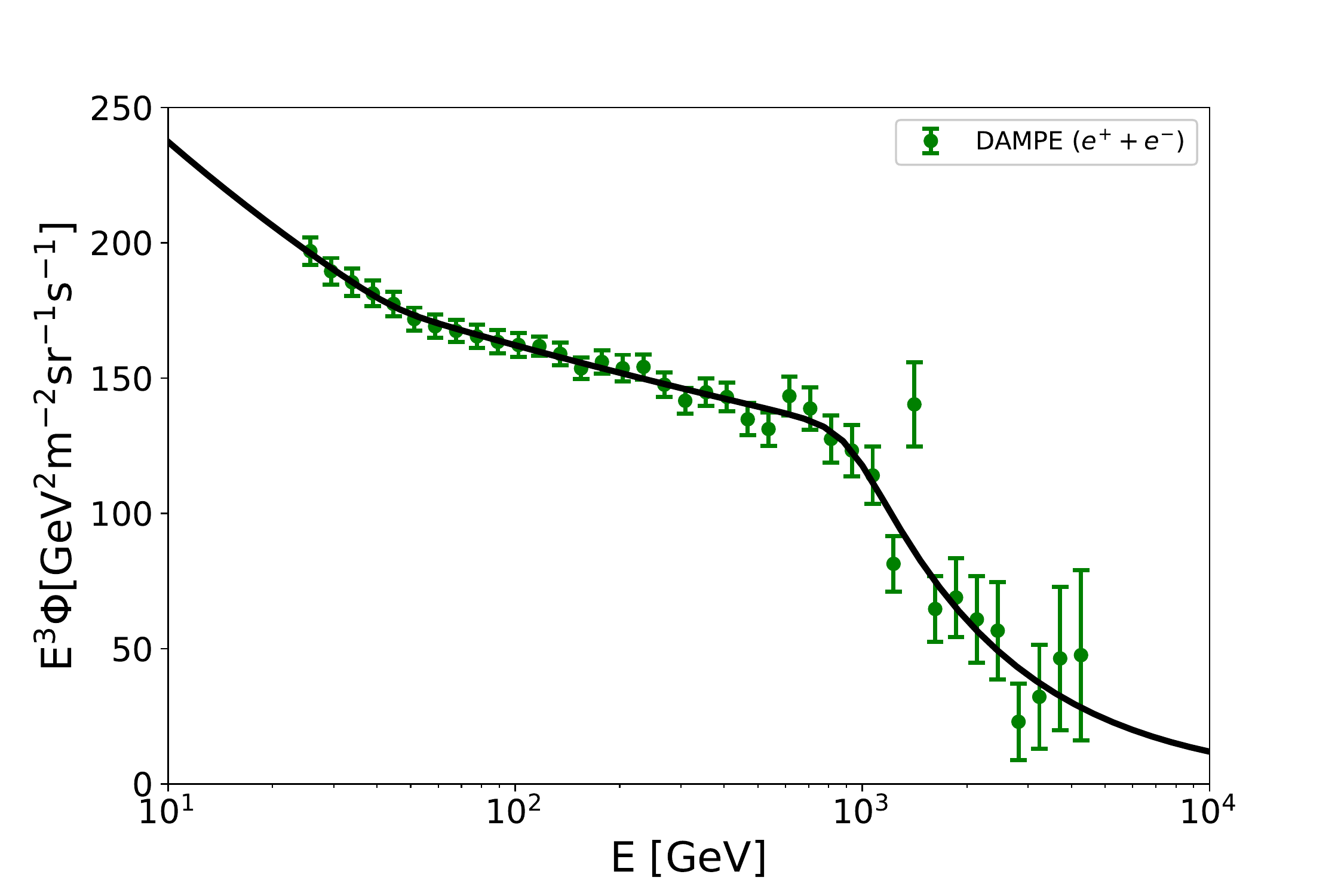}
    \includegraphics[width=0.46\columnwidth]{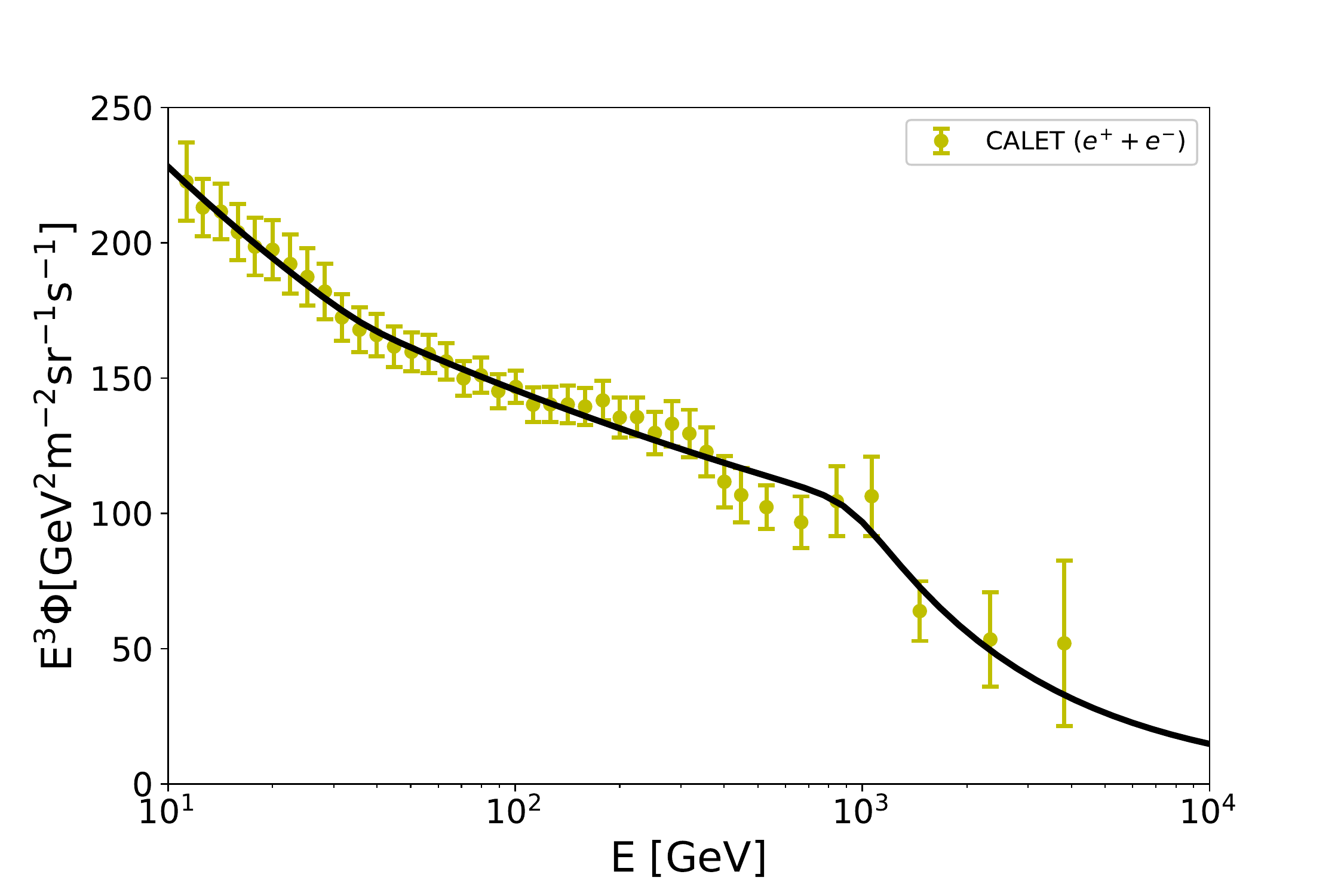}
    \includegraphics[width=0.46\columnwidth]{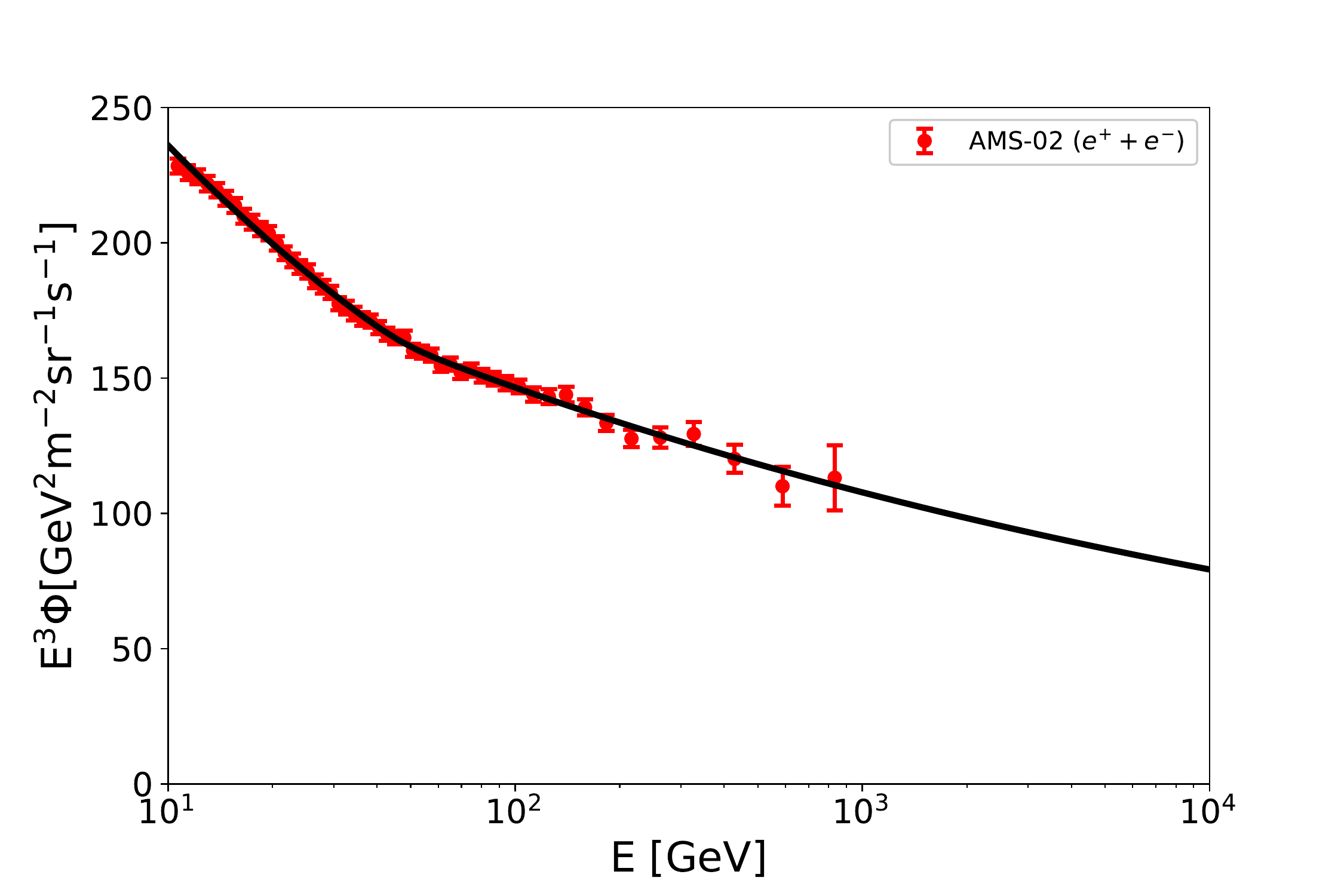}
   \includegraphics[width=0.46\columnwidth]{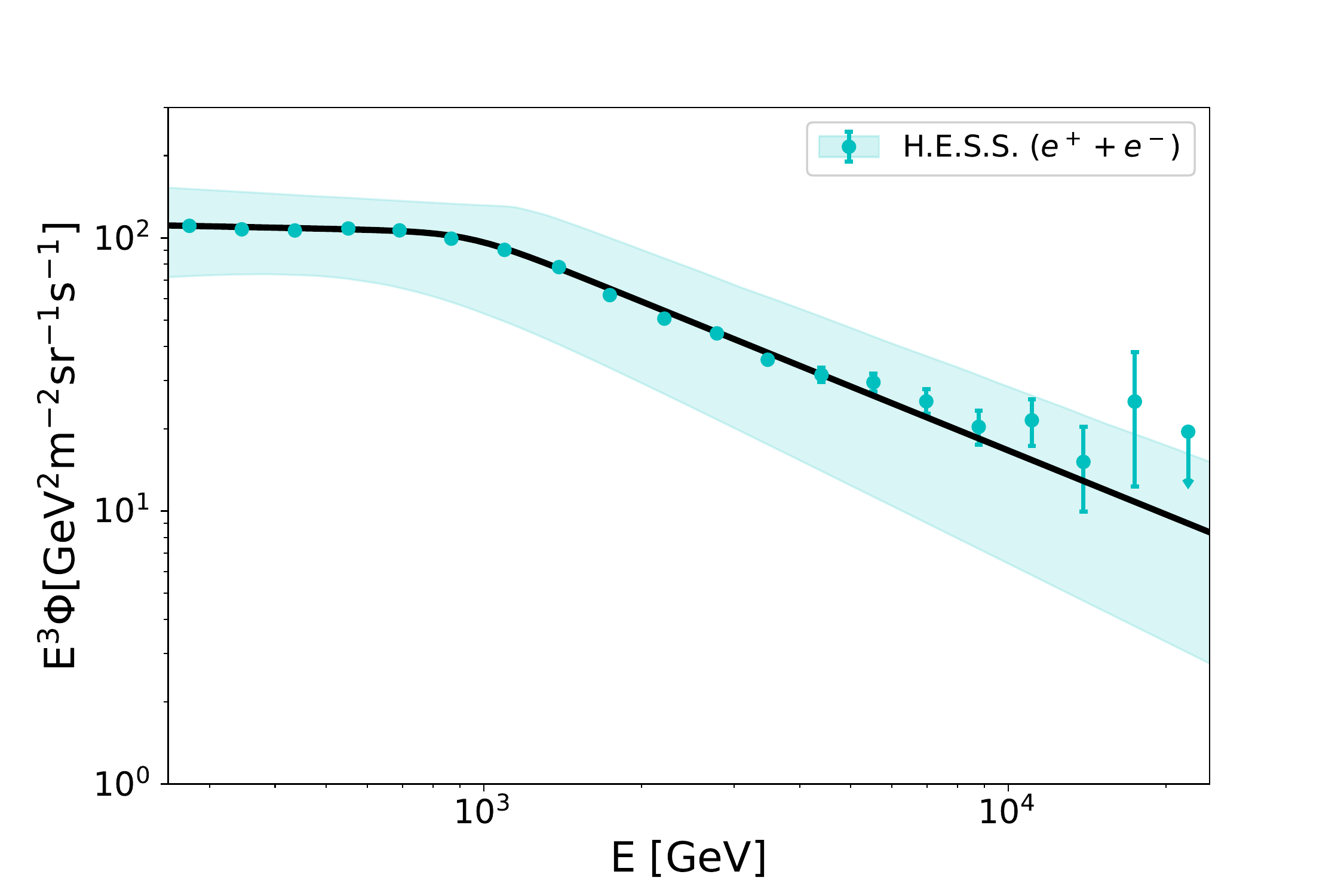}
   \includegraphics[width=0.46\columnwidth]{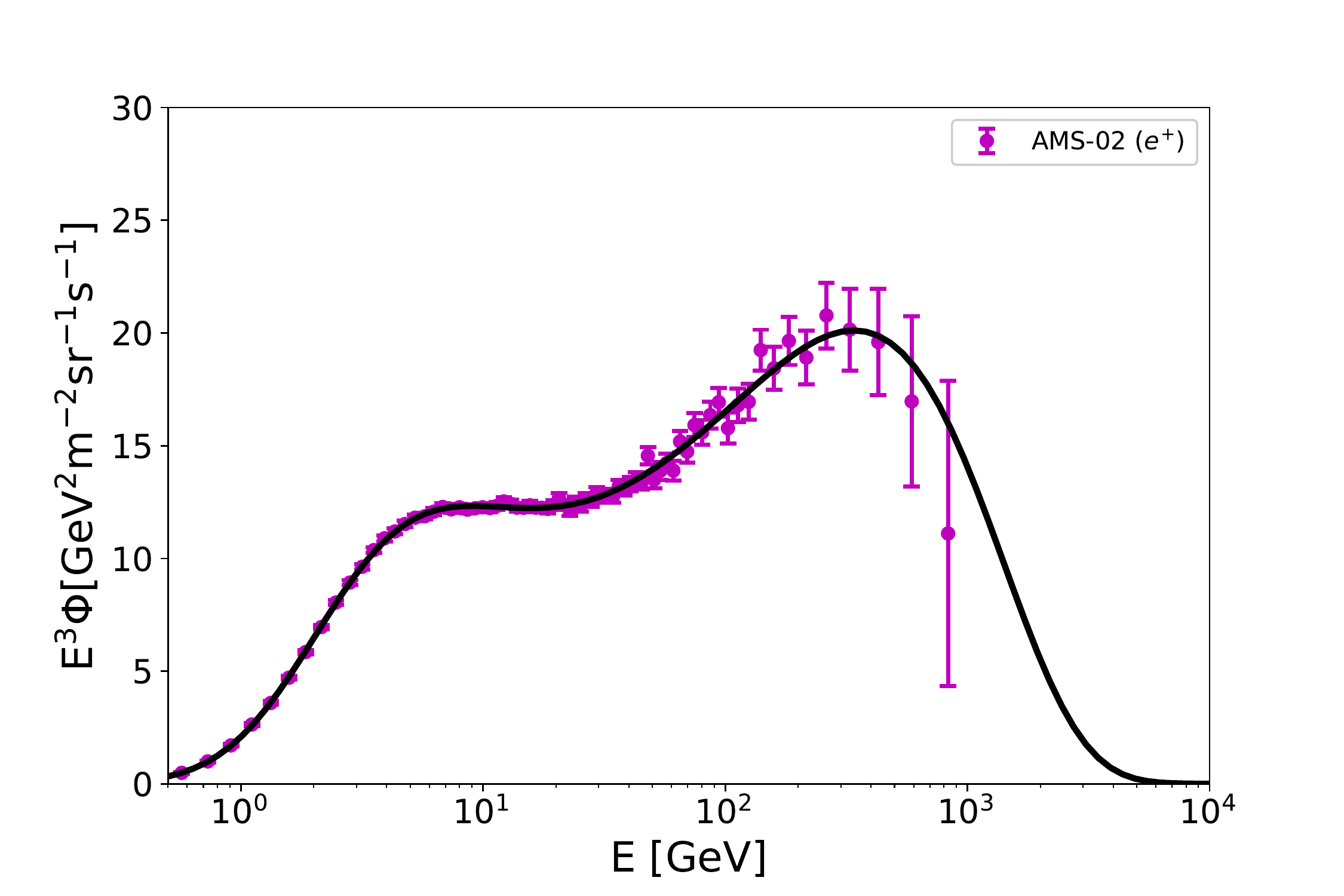}
  \end{center}
  \caption{Best-fit fluxes from fitting to the CRE or CR positron spectra.
  The top four panels are from the fits of \Eq{eq:smooth_pl} to the CRE data 
  with energies above 10 GeV from Fermi-LAT \cite{Abdollahi:2017nat}, DAMPE \cite{Ambrosi:2017wek}, 
  CALET \cite{Adriani:2018ktz}, and AMS-02 \cite{Aguilar:2019ksn} respectively. 
  The left bottom panel is from the fit of \Eq{eq:hess_spl} to the CRE data of 
  H.E.S.S. from \cite{HESSICRC17}. The right bottom panel is from the fit of \Eq{eq:dps} 
  to the CR positron data of AMS-02 from \cite{Aguilar:2019owu}.}
  \label{fig:eachCREfit}
\end{figure}

To evaluate the systematic uncertainties on the fluxes of CRE and CR positron 
due to the energy scale uncertainties with \Eq{eq:flux_uncertainty}, 
one needs to know the flux $\Phi(E)$ and its derivative $\Phi^{\prime}(E)$ first.
In this work, the flux $\Phi(E)$ is approximated by a smooth curve with 
parameters determined through fitting to the flux data. The derivative 
$\Phi^{\prime}(E)$  can then be obtained straightforwardly from $\Phi(E)$. The 
CRE data from Fermi-LAT \cite{Abdollahi:2017nat}, DAMPE \cite{Ambrosi:2017wek}, 
CALET \cite{Adriani:2018ktz}, and AMS-02 \cite{Aguilar:2019ksn} are fitted 
with a smoothly broken power-law curve given by \Eq{eq:smooth_pl}. 
The data with energy above 10 GeV are considered. The best-fit parameters 
and the goodness-of-fit of each individual fit are summarized 
in \Tab{tab:eachCREfit}. From \Tab{tab:eachCREfit}, one can 
see that the smoothly broken power-law model is a good approximation to 
the measured spectrum as the $\chi^{2} / \mathrm{d.o.f.}$ of each individual 
fit is less than 1.
For the CRE data of H.E.S.S. \cite{HESSICRC17}, we adopt the parameterization 
reported in the International Cosmic Ray Conference \cite{HESSICRC17}:
\begin{equation}\label{eq:hess_spl}
  E^{3} \frac{\mathrm{d} N}{\mathrm{d} E}=N_{0}
  \left(\frac{E}{(1 \mathrm{TeV})}\right)^{3-\Gamma_{1}}
  \left(1+\left(\frac{E}{E_{b}}\right)^{\frac{1}{\alpha}}\right)^{-\left(\Gamma_{2}-\Gamma_{1}\right) \alpha}.
\end{equation}
\Tab{tab:hessfit} lists the best-fit parameters from \cite{HESSICRC17}, which
are obtained through fitting to the H.E.S.S. CRE data.
The latest CR positron data from AMS-02 is well described by the 
minimal model \cite{Aguilar:2013qda,Accardo:2014lma,Cavasonza:2016qem,
Aguilar:2019owu}, in which the positron flux is parametrized as the sum of a 
diffuse term and a source term
\begin{equation} \label{eq:dps}
  \Phi_{e^{+}}(E)= \frac{E^{2}}{\hat{E}^{2}}\left[C_{d}\left(\hat{E} / E_{1}\right)^{\gamma_{d}}
  +C_{s}\left(\hat{E} / E_{2}\right)^{\gamma_{s}} \exp \left(-\hat{E} / E_{s}\right) \right]    ,
\end{equation}
where $\hat{E}=E+\phi_{e^{+}}$ is the energy of particles in the interstellar 
space and $\phi_{e^{+}}$ is the effective solar potential. 
\Tab{tab:positronfit} lists the best-fit parameters from \cite{Aguilar:2019owu},
which are obtained by fitting to the latest AMS-02 positron data with the 
minimal model. In \Fig{fig:eachCREfit}, we illustrate the comparisons of the 
best-fitting curves and the measured spectra for all the experimental data 
described above.

\begin{table}[!t]
  \begin{center}
  \scalebox{0.88}{
  \begin{threeparttable}
  \begin{tabular}{L{1.6cm} | C{2.3cm} C{2.3cm} C{2.3cm} C{2.2cm} C{1.9cm} C{2.7cm} | C{2.0cm} }    \hline \hline
  \textbf{CRE} & $\Phi_{0}$ & $\gamma_{1}$ & $\gamma_{2}$ & $\gamma_{3}$ & $E_{\mathrm{br1}}$ & $E_{\mathrm{br2}}$ & $\bm{\chi^{2} / \mathrm{d.o.f.}}$ \\ \hline
  FERMI\tnote{$\dagger$}     & 5.40 $\pm$ 0.05 & 3.21 $\pm$ 0.00   & 3.06 $\pm$ 0.01   & 3.27 $\pm$ 0.13   & 56.1 $\pm$ 3.4    & 822.6 $\pm$ 274.7 & $\bm{3.7/40}$ \\
  DAMPE                      & 5.42 $\pm$ 0.05 & 3.20 $\pm$ 0.08   & 3.09 $\pm$ 0.01   & 4.01 $\pm$ 0.18   & 45.9 $\pm$ 16.5   & 925.2 $\pm$ 92.4  & $\bm{25.6/32}$ \\
  CALET                      & 4.59 $\pm$ 0.05 & 3.23 $\pm$ 0.03   & 3.15 $\pm$ 0.01   & 3.83 $\pm$ 0.29   & 37.0 $\pm$ 13.7   & 945.6 $\pm$ 200.7 & $\bm{13.3/34}$ \\
  AMS-02                     & 4.69 $\pm$ 0.03 & 3.242 $\pm$ 0.004 & 3.133 $\pm$ 0.005 & -                 & 47.9 $\pm$ 2.9    & -                 & $\bm{15.8/46}$ \\
  \hline\hline
  \end{tabular}
  \begin{tablenotes}
  \footnotesize
  \item[$\dagger$] Considered the LAT energy reconstruction uncertainty.
  \end{tablenotes}
  \end{threeparttable}}
  \end{center}
  \caption{The best-fit parameters corresponding to the fit of \Eq{eq:smooth_pl} to 
  the CRE data with energy above 10 GeV from Fermi-LAT \cite{Abdollahi:2017nat}, 
  DAMPE \cite{Ambrosi:2017wek}, CALET \cite{Adriani:2018ktz}, and AMS-02 
  \cite{Aguilar:2019ksn}. The reduced $\chi^2$ of each fit is also listed. 
  $\Phi_{0}$ is in units of $10^{-6}~\mathrm{m}^{-2} \mathrm{sr}^{-1} \mathrm{s}^{-1} \mathrm{GeV}^{-1}$.
  $E_{\mathrm{br1}}$ and $E_{\mathrm{br2}}$ are in units of $\mathrm{GeV}$.}
  \label{tab:eachCREfit}
\end{table}

\begin{table}[!t]
  \begin{tabular}{C{2.7cm} | C{2.6cm} C{2.5cm} C{2.5cm} C{2.5cm} C{2.6cm} }    \hline \hline
  \textbf{CRE} & $N_{0}$ & $\Gamma_{1}$ & $\Gamma_{2}$ & $E_{b}$ & $\alpha$ \\
  \hline
  H.E.S.S.  & 105 $\pm$ 1 & 3.04 $\pm$ 0.01 & 3.78 $\pm$ 0.02 & 0.94 $\pm$ 0.02 & 0.12 $\pm$ 0.01 \\
  \hline\hline
  \end{tabular}
  \caption{The best-fit parameters corresponding to the fit of \Eq{eq:hess_spl}
  to the CRE data of H.E.S.S. from \cite{HESSICRC17}. $N_{0}$ is in units of 
  $\mathrm{m}^{-2} \mathrm{sr}^{-1} \mathrm{s}^{-1} \mathrm{GeV}^{2}$. $E_{b}$ 
  is in units of $\mathrm{TeV}$.}
  \label{tab:hessfit}
\end{table}

\begin{table}[!t]
  \begin{tabular}{C{2.2cm} | C{2.2cm} C{2.2cm} C{2.2cm} C{2.2cm} C{2.2cm} C{2.2cm}}    \hline \hline
  \textbf{Positron} & $1 / E_{s}$     & $C_{s}$ & $\gamma_{s}$ & $C_{d}$ & $\gamma_{d}$ & $\varphi_{e^+}$ \\ \hline
  AMS-02   & 1.23 $\pm$ 0.34 & 6.80 $\pm$ 0.15 & -2.58 $\pm$ 0.05 & 6.51 $\pm$ 0.14 & -4.07 $\pm$ 0.06 & 1.10 $\pm$ 0.03 \\
  \hline \hline
  \end{tabular}
  \caption{The best-fit parameters corresponding to the fit of \Eq{eq:dps} to 
  the CR positron data of AMS-02 from \cite{Aguilar:2019owu}. $E_{s}$ and 
  $\varphi_{e^{+}}$ are in units of $\mathrm{TeV}$ and $\mathrm{GeV}$, respectively.
  $C_{s}$ and $C_{d}$ are in units of $10^{-5}$ and $10^{-2}$ 
  $[\mathrm{m}^{-2} \mathrm{sr}^{-1} \mathrm{s}^{-1} \mathrm{GeV}^{-1}]$, respectively.}
  \label{tab:positronfit}
\end{table}

\begin{figure}[t]
  \begin{center}
    \includegraphics[width=\columnwidth]{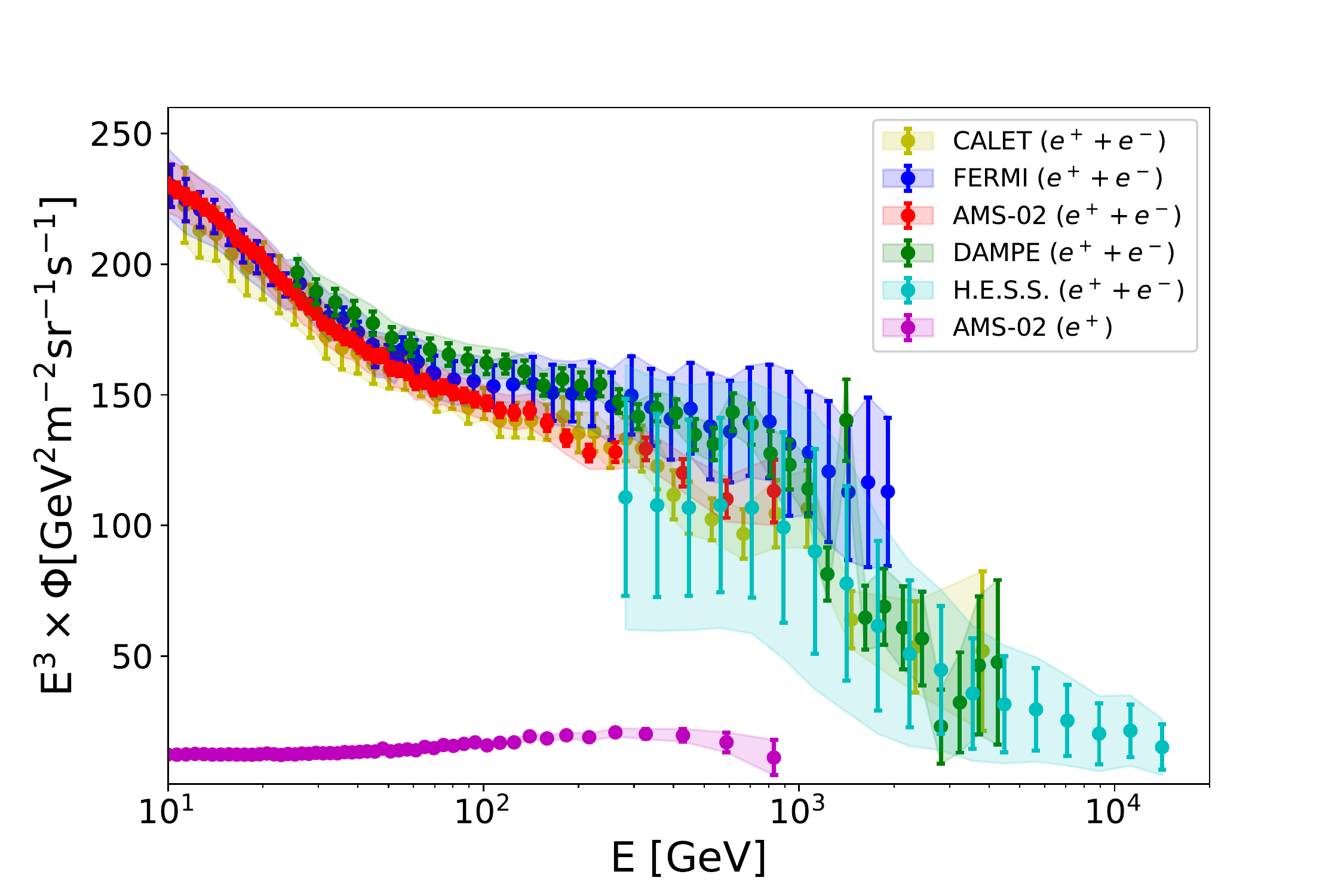}
  \end{center}
\caption{The CRE spectra of Fermi-LAT \cite{Abdollahi:2017nat} (blue), 
DAMPE \cite{Ambrosi:2017wek} (green), CALET \cite{Adriani:2018ktz} (yellow),
AMS-02 \cite{Aguilar:2019ksn} (red), and H.E.S.S. \cite{HESSICRC17} (cyan) and 
the CR positron spectrum of AMS-02
\cite{Aguilar:2019owu} (purple) rescaled by $E^3$. The uncertainty bands and 
the error bars represent the total uncertainties (quadratic sum of the 
statistical and systematic uncertainties) with and without including the 
energy scale uncertainties, respectively.}
\label{fig:fitting_datas}
\end{figure}

Given $\Phi(E)$, $\Phi^{\prime}(E)$ and the energy scale uncertainty $\delta_s$
(summarized in \Sec{sec:Experimental_data}), 
the systematic uncertainty on the flux due to the energy scale uncertainty
can be obtained straightforward from \Eq{eq:flux_uncertainty}. In this work, 
we calculate this part systematic uncertainty for the CRE data from Fermi-LAT,
DAMPE, CALET, AMS-02, and H.E.S.S. and for the CR positron data from AMS-02. 
The total uncertainties (quadratic sum of the statistical and systematic 
uncertainties) in the data with and without including this part uncertainty 
for each experiment are summarized in \Tab{tab:ESU_FDC} and 
\Tab{tab:ESU_AMS}, and shown in \Fig{fig:fitting_datas}.

\section{Statistical framework}\label{app:bayesian}
Bayesian inference method provides a consistent approach both to the estimation 
of a set of parameters $\mathbf{\Theta}$ in a model (or hypothesis) $H$ for the 
data $\mathbf{D}$ and the evaluation of the relative advantage of different models
for the data. This approach evaluates the posterior probability 
distribution function (PDF) for the parameters of interest in a given model
through Bayes' theorem, which states that
\begin{equation}
  P(\mathbf{\Theta}|\mathbf{D}, H) =
  \frac{P(\mathbf{D}|\,\mathbf{\Theta},H) P(\mathbf{\Theta}|H)}{P(\mathbf{D}|H)},
\end{equation}
where $P(\mathbf{\Theta}|\mathbf{D}, H)$ is the posterior PDF, 
$P(\mathbf{D}|\mathbf{\Theta}, H) \equiv \mathcal {L}(\mathbf{\Theta})$ is 
the likelihood function which contains the information provided by the data, 
and $P(\mathbf{\Theta}|H) \equiv \pi(\mathbf{\Theta})$ is the prior 
PDF of the parameters which encompasses our state of knowledge on the
values of the parameters before the observation of the data. The quantity 
$P(\mathbf{D}|H) \equiv \mathcal{Z}$ is the Bayesian evidence which is
obtained by integrating the product of the likelihood and the prior over
the whole volume of the parameter space
\begin{equation}
  \mathcal{Z} = \int{\mathcal{L}(\mathbf{\Theta})\pi(\mathbf{\Theta})}d\mathbf{\Theta}    .
\end{equation}
Since the Bayesian evidence is independent of the parameter values 
$\mathbf{\Theta}$, it is usually ignored in parameter estimation problems and 
the posterior inferences are obtained by exploring the unnormalized 
posterior using standard Markov chain Monte Carlo sampling methods.

In contrast to parameter estimation problems, the Bayesian evidence takes the 
central role in model selection. In order to select between two models
$H_{i}$ and $H_{j}$, one needs to compare their respective posterior PDFs 
given the observed dataset $\mathbf{D}$, as follows:
\begin{equation}
  \frac{P(H_{i}|\mathbf{D})}{P(H_{j}|\mathbf{D})}
  =\frac{P(\mathbf{D}|H_{i})P(H_{i})/P(\mathbf{D})}{P(\mathbf{D}|H_{j})P(H_{j})/P(\mathbf{D})}
  =\frac{\mathcal{Z}_i}{\mathcal{Z}_j}\frac{P(H_{i})}{P(H_{j})},
\end{equation}
where $P(H_{i})/P(H_{j})$ is the prior probability ratio for the two models, 
and is usually assumed to be unity. The evidence ratio 
\begin{equation}
  K_{ij}\equiv\mathcal{Z}_i/\mathcal{Z}_j
\end{equation}
is the so-called Bayes factor between the two models. \Tab{tab:bf} lists the 
categories for interpreting the Bayes factor, which is given by Kass and 
Raftery (1995) \cite{doi:10.1080/01621459.1995.10476572}.

In this work, we take the prior PDF as a uniform distribution
\begin{equation}
  \pi\left(\theta_{i}\right) \propto\left\{\begin{array}{ll}
  \frac{1}{\theta_{i,\max} - \theta_{i,\min}}, & \text { for } \theta_{i, \min }\leq\theta_{i}\leq\theta_{i, \max } \\
  0, & \text { otherwise }
  \end{array}\right.  ,
\end{equation}
and the likelihood function as Gaussian form
\begin{equation}
  \mathcal {L}(\mathbf{\Theta}) = 
  \prod_{i=1} \frac{1}{\sqrt{2\pi\sigma_{\mathrm{exp}, i}^2}}\exp \left[
  - \frac{\left(\Phi_{\mathrm{th}, i}(\mathbf{\Theta}) - \Phi_{\mathrm{exp}, i}\right)^2}{2\sigma_{\mathrm{exp}, i}^2} 
  \right]    ,
\end{equation}
where $\Phi_{\mathrm{th}, i}(\mathbf{\Theta})$ is the $i$-th theoretical 
predicted value from the model which depends on the parameters $\mathbf{\Theta}$, 
and $\Phi_{\mathrm{exp}, i}$ is the one measured by the experiment with 
uncertainty $\sigma_{\mathrm{exp}, i}$. 
We estimate the parameters of our models and evaluate the Bayesian evidence 
for each model by using the public code {\tt MultiNest} \cite{Feroz:2007kg}, 
which is a highly efficient implementation of the nested sampling technique 
and is fully parallelized. More details of the algorithm can be found 
in \cite{Feroz:2007kg,Feroz:2008xx,Feroz:2013hea}. Here we summarize the main 
settings of {\tt MultiNest} used in this work. 
The number of live points, which influences the accuracy of evidence 
estimation and convergence rate of the algorithm is taken to be 1000, which 
is sufficient enough. 
The sampling efficiency is taken to be 0.3, which is recommended for the
evidence evaluation. 
Lastly, we choose a tolerance of 0.1, which controls the precision to be 
achieved on the evidence.

\begin{table}[t]
  \begin{tabular}{R{1.5cm} C{2.5cm} L{6.5cm}}       \hline\hline
     $2\ln K$ &     $K$       & Strength of evidence \\   \hline
    0 to 2    &   1 to 3      & not worth more than a bare mention    \\
    2 to 6    &   3 to 20     & positive  \\
    6 to 10   &   20 to 150   & strong    \\
    $> 10$    &   $> 150$     & very strong    \\   \hline \hline
  \end{tabular}
  \caption{Interpretation of Bayes factor $K$ from \cite{doi:10.1080/01621459.1995.10476572}.}
  \label{tab:bf}
\end{table}

\clearpage

\clearpage
\begin{table}[h]
  \scriptsize
  \begin{tabular}{C{1.1cm} R{1.2cm} C{.7cm} L{1.9cm} | C{1.1cm} R{1.2cm} C{.7cm} L{1.9cm} | C{1.1cm} R{1.2cm} C{.7cm} L{1.9cm}}  \hline \hline
  \multicolumn{4}{c|}{\textbf{FERMI}}  & \multicolumn{4}{c|}{\textbf{DAMPE}} & \multicolumn{4}{c}{\textbf{CALET}} \\ \hline
  E [GeV] & $\Phi_{e^+ + e^-}$ & $\sigma_{\mathrm{tot}}$ & $\sigma_{\mathrm{tot}}^{\prime}$ & E [GeV] & $\Phi_{e^+ + e^-}$ & $\sigma_{\mathrm{tot}}$ & $\sigma_{\mathrm{tot}}^{\prime}$ & E [GeV] & $\Phi_{e^+ + e^-}$ & $\sigma_{\mathrm{tot}}$ & $\sigma_{\mathrm{tot}}^{\prime}$  \\ \hline
  10.2    & $(2.148$ & $0.076$ & $0.122) \times 10^{-1}$  & 25.7    & $(1.160$ & $0.030$ & $0.044) \times 10^{-2}$  & 11.3    & $(1.543$ & $0.100$ & $0.105) \times 10^{-1}$  \\
  11.4    & $(1.525$ & $0.056$ & $0.087) \times 10^{-1}$  & 29.5    & $(7.380$ & $0.191$ & $0.280) \times 10^{-3}$  & 12.6    & $(1.065$ & $0.053$ & $0.057) \times 10^{-1}$  \\
  12.6    & $(1.094$ & $0.033$ & $0.059) \times 10^{-1}$  & 33.9    & $(4.760$ & $0.132$ & $0.186) \times 10^{-3}$  & 14.2    & $(7.388$ & $0.358$ & $0.388) \times 10^{-2}$  \\
  14.0    & $(7.903$ & $0.232$ & $0.420) \times 10^{-2}$  & 38.9    & $(3.080$ & $0.081$ & $0.117) \times 10^{-3}$  & 15.9    & $(5.073$ & $0.257$ & $0.277) \times 10^{-2}$  \\
  15.6    & $(5.671$ & $0.176$ & $0.307) \times 10^{-2}$  & 44.6    & $(2.000$ & $0.051$ & $0.075) \times 10^{-3}$  & 17.8    & $(3.521$ & $0.188$ & $0.201) \times 10^{-2}$  \\
  17.3    & $(4.016$ & $0.123$ & $0.216) \times 10^{-2}$  & 51.2    & $(1.280$ & $0.032$ & $0.047) \times 10^{-3}$  & 20.0    & $(2.468$ & $0.137$ & $0.146) \times 10^{-2}$  \\
  19.2    & $(2.882$ & $0.088$ & $0.155) \times 10^{-2}$  & 58.8    & $(8.320$ & $0.214$ & $0.307) \times 10^{-4}$  & 22.5    & $(1.687$ & $0.096$ & $0.102) \times 10^{-2}$  \\
  21.2    & $(2.062$ & $0.060$ & $0.109) \times 10^{-2}$  & 67.6    & $(5.420$ & $0.133$ & $0.196) \times 10^{-4}$  & 25.2    & $(1.171$ & $0.066$ & $0.070) \times 10^{-2}$  \\
  23.6    & $(1.468$ & $0.034$ & $0.073) \times 10^{-2}$  & 77.6    & $(3.540$ & $0.092$ & $0.131) \times 10^{-4}$  & 28.3    & $(8.029$ & $0.452$ & $0.480) \times 10^{-3}$  \\
  26.2    & $(1.075$ & $0.021$ & $0.052) \times 10^{-2}$  & 89.1    & $(2.310$ & $0.061$ & $0.086) \times 10^{-4}$  & 31.7    & $(5.413$ & $0.272$ & $0.293) \times 10^{-3}$  \\
  29.1    & $(7.542$ & $0.161$ & $0.370) \times 10^{-3}$  & 102.2   & $(1.520$ & $0.041$ & $0.058) \times 10^{-4}$  & 35.6    & $(3.721$ & $0.183$ & $0.197) \times 10^{-3}$  \\
  32.3    & $(5.328$ & $0.125$ & $0.267) \times 10^{-3}$  & 117.4   & $(1.000$ & $0.022$ & $0.035) \times 10^{-4}$  & 39.9    & $(2.612$ & $0.124$ & $0.134) \times 10^{-3}$  \\
  35.9    & $(3.871$ & $0.092$ & $0.194) \times 10^{-3}$  & 134.8   & $(6.490$ & $0.171$ & $0.242) \times 10^{-5}$  & 44.8    & $(1.798$ & $0.084$ & $0.091) \times 10^{-3}$  \\
  40.0    & $(2.725$ & $0.064$ & $0.137) \times 10^{-3}$  & 154.8   & $(4.140$ & $0.108$ & $0.153) \times 10^{-5}$  & 50.3    & $(1.255$ & $0.057$ & $0.062) \times 10^{-3}$  \\
  44.5    & $(1.920$ & $0.050$ & $0.098) \times 10^{-3}$  & 177.7   & $(2.780$ & $0.076$ & $0.106) \times 10^{-5}$  & 56.4    & $(8.863$ & $0.393$ & $0.429) \times 10^{-4}$  \\
  45.3    & $(1.779$ & $0.052$ & $0.094) \times 10^{-3}$  & 204.0   & $(1.810$ & $0.058$ & $0.075) \times 10^{-5}$  & 63.3    & $(6.157$ & $0.267$ & $0.292) \times 10^{-4}$  \\
  49.6    & $(1.354$ & $0.036$ & $0.069) \times 10^{-3}$  & 234.2   & $(1.200$ & $0.036$ & $0.048) \times 10^{-5}$  & 71.0    & $(4.188$ & $0.180$ & $0.197) \times 10^{-4}$  \\
  52.3    & $(1.137$ & $0.036$ & $0.061) \times 10^{-3}$  & 268.9   & $(7.590$ & $0.236$ & $0.309) \times 10^{-6}$  & 79.7    & $(2.984$ & $0.128$ & $0.141) \times 10^{-4}$  \\
  55.3    & $(9.886$ & $0.272$ & $0.503) \times 10^{-4}$  & 308.8   & $(4.810$ & $0.163$ & $0.206) \times 10^{-6}$  & 89.4    & $(2.032$ & $0.088$ & $0.096) \times 10^{-4}$  \\
  60.4    & $(7.421$ & $0.329$ & $0.454) \times 10^{-4}$  & 354.5   & $(3.250$ & $0.113$ & $0.142) \times 10^{-6}$  & 100.4   & $(1.450$ & $0.059$ & $0.065) \times 10^{-4}$  \\
  61.8    & $(6.891$ & $0.249$ & $0.382) \times 10^{-4}$  & 407.1   & $(2.120$ & $0.078$ & $0.096) \times 10^{-6}$  & 112.6   & $(9.820$ & $0.443$ & $0.482) \times 10^{-5}$  \\
  69.0    & $(4.818$ & $0.244$ & $0.315) \times 10^{-4}$  & 467.4   & $(1.320$ & $0.058$ & $0.068) \times 10^{-6}$  & 126.2   & $(6.980$ & $0.326$ & $0.353) \times 10^{-5}$  \\
  69.8    & $(4.655$ & $0.197$ & $0.276) \times 10^{-4}$  & 536.6   & $(8.490$ & $0.400$ & $0.458) \times 10^{-7}$  & 141.7   & $(4.930$ & $0.242$ & $0.260) \times 10^{-5}$  \\
  80.6    & $(2.978$ & $0.134$ & $0.182) \times 10^{-4}$  & 616.1   & $(6.130$ & $0.309$ & $0.349) \times 10^{-7}$  & 159.0   & $(3.470$ & $0.171$ & $0.183) \times 10^{-5}$  \\
  93.1    & $(1.927$ & $0.095$ & $0.124) \times 10^{-4}$  & 707.4   & $(3.920$ & $0.224$ & $0.248) \times 10^{-7}$  & 178.8   & $(2.480$ & $0.126$ & $0.135) \times 10^{-5}$  \\
  107.5   & $(1.235$ & $0.064$ & $0.082) \times 10^{-4}$  & 812.2   & $(2.380$ & $0.162$ & $0.176) \times 10^{-7}$  & 200.1   & $(1.690$ & $0.092$ & $0.098) \times 10^{-5}$  \\
  124.1   & $(8.059$ & $0.458$ & $0.565) \times 10^{-5}$  & 932.5   & $(1.520$ & $0.117$ & $0.127) \times 10^{-7}$  & 224.4   & $(1.200$ & $0.064$ & $0.068) \times 10^{-5}$  \\
  143.3   & $(5.242$ & $0.347$ & $0.409) \times 10^{-5}$  & 1070.7  & $(9.290$ & $0.863$ & $0.925) \times 10^{-8}$  & 252.5   & $(8.060$ & $0.488$ & $0.512) \times 10^{-6}$  \\
  165.5   & $(3.329$ & $0.238$ & $0.275) \times 10^{-5}$  & 1229.3  & $(4.380$ & $0.548$ & $0.572) \times 10^{-8}$  & 282.9   & $(5.880$ & $0.369$ & $0.386) \times 10^{-6}$  \\
  191.1   & $(2.155$ & $0.153$ & $0.177) \times 10^{-5}$  & 1411.4  & $(4.990$ & $0.557$ & $0.588) \times 10^{-8}$  & 317.4   & $(4.050$ & $0.275$ & $0.286) \times 10^{-6}$  \\
  220.7   & $(1.398$ & $0.113$ & $0.127) \times 10^{-5}$  & 1620.5  & $(1.520$ & $0.286$ & $0.292) \times 10^{-8}$  & 355.6   & $(2.730$ & $0.200$ & $0.207) \times 10^{-6}$  \\
  254.8   & $(8.799$ & $0.780$ & $0.860) \times 10^{-6}$  & 1860.6  & $(1.070$ & $0.226$ & $0.229) \times 10^{-8}$  & 400.4   & $(1.740$ & $0.148$ & $0.151) \times 10^{-6}$  \\
  294.3   & $(5.876$ & $0.585$ & $0.633) \times 10^{-6}$  & 2136.3  & $(6.240$ & $1.638$ & $1.655) \times 10^{-9}$  & 447.7   & $(1.190$ & $0.112$ & $0.114) \times 10^{-6}$  \\
  339.8   & $(3.702$ & $0.376$ & $0.406) \times 10^{-6}$  & 2452.8  & $(3.840$ & $1.218$ & $1.227) \times 10^{-9}$  & 529.3   & $(6.900$ & $0.541$ & $0.558) \times 10^{-7}$  \\
  392.4   & $(2.330$ & $0.258$ & $0.275) \times 10^{-6}$  & 2816.1  & $(1.030$ & $0.634$ & $0.635) \times 10^{-9}$  & 666.3   & $(3.270$ & $0.323$ & $0.330) \times 10^{-7}$  \\
  453.2   & $(1.556$ & $0.187$ & $0.198) \times 10^{-6}$  & 3233.4  & $(9.530$ & $5.683$ & $5.695) \times 10^{-10}$ & 843.7   & $(1.740$ & $0.216$ & $0.219) \times 10^{-7}$  \\
  523.3   & $(9.622$ & $1.414$ & $1.468) \times 10^{-7}$  & 3712.4  & $(9.070$ & $5.178$ & $5.189) \times 10^{-10}$ & 1063.6  & $(8.840$ & $1.220$ & $1.239) \times 10^{-8}$  \\
  604.3   & $(6.159$ & $0.884$ & $0.920) \times 10^{-7}$  & 4262.4  & $(6.150$ & $4.064$ & $4.071) \times 10^{-10}$ & 1463.2  & $(2.040$ & $0.352$ & $0.356) \times 10^{-8}$  \\
  697.8   & $(4.112$ & $0.607$ & $0.631) \times 10^{-7}$  &  &      &  &                                            & 2336.2  & $(4.190$ & $1.370$ & $1.374) \times 10^{-9}$  \\
  805.8   & $(2.671$ & $0.415$ & $0.431) \times 10^{-7}$  &  &      &  &                                            & 3815.3  & $(9.360$ & $5.496$ & $5.501) \times 10^{-10}$ \\
  930.6   & $(1.628$ & $0.341$ & $0.349) \times 10^{-7}$  &  &      &  &                                            &  &      &  &                                            \\
  1074.6  & $(1.031$ & $0.188$ & $0.193) \times 10^{-7}$  &  &      &  &                                            &  &      &  &                                            \\
  1240.9  & $(6.314$ & $1.411$ & $1.439) \times 10^{-8}$  &  &      &  &                                            &  &      &  &                                            \\
  1433.0  & $(3.833$ & $0.882$ & $0.899) \times 10^{-8}$  &  &      &  &                                            &  &      &  &                                            \\
  1654.8  & $(2.571$ & $0.716$ & $0.725) \times 10^{-8}$  &  &      &  &                                            &  &      &  &                                            \\
  1911.0  & $(1.618$ & $0.406$ & $0.413) \times 10^{-8}$  &  &      &  &                                            &  &      &  &                                            \\
  \hline \hline
  \end{tabular}
  \caption{The CRE spectra of Fermi-LAT \cite{Abdollahi:2017nat}, DAMPE 
  \cite{Ambrosi:2017wek}, and CALET \cite{Adriani:2018ktz}. The parameters
  $\sigma_{\mathrm{tot}}^{\prime}$
  and $\sigma_{\mathrm{tot}}$ represent the total uncertainties (quadratic sum
  of the statistical and systematic uncertainties) with and without including 
  the energy scale uncertainties, respectively. The flux $\Phi_{e^+ + e^-}$ is in units of 
  $\mathrm{m}^{-2} \mathrm{sr}^{-1} \mathrm{s}^{-1} \mathrm{GeV}^{-1}$.}
  \label{tab:ESU_FDC}
\end{table}
  
\clearpage
\begin{table}[h]
  \scriptsize
  \begin{tabular}{C{1.1cm} R{1.2cm} C{.7cm} L{2.2cm} | C{1.7cm} R{1.2cm} C{.7cm} L{2.0cm} R{1.2cm} C{.7cm} L{2.1cm} }  \hline \hline
    \multicolumn{4}{c|}{\textbf{H.E.S.S.}} & \multicolumn{7}{c}{\textbf{AMS-02}} \\ \hline
    E [GeV] & $\Phi_{e^+ + e^-}$ & $\sigma_{\mathrm{tot}}$ & $\sigma_{\mathrm{tot}}^{\prime}$ & E [GeV] & $\Phi_{e^+ + e^-}$ & $\sigma_{\mathrm{tot}}$ & $\sigma_{\mathrm{tot}}^{\prime}$ & $\Phi_{e^+}$ & $\sigma_{\mathrm{tot}}$ & $\sigma_{\mathrm{tot}}^{\prime}$ \\ \hline
    281.7   & $(4.955$ & $1.692$ & $2.272) \times 10^{-6}$  & 10.67   & $(1.880$ & $0.023$ & $0.083) \times 10^{-1}$  & $(1.007$ & $0.014$ & $0.041) \times 10^{-2}$  \\
    355.0   & $(2.409$ & $0.785$ & $1.077) \times 10^{-6}$  & 11.41   & $(1.521$ & $0.019$ & $0.067) \times 10^{-1}$  & $(8.302$ & $0.121$ & $0.340) \times 10^{-3}$  \\
    447.3   & $(1.193$ & $0.376$ & $0.524) \times 10^{-6}$  & 12.19   & $(1.239$ & $0.015$ & $0.055) \times 10^{-1}$  & $(6.918$ & $0.102$ & $0.284) \times 10^{-3}$  \\
    563.6   & $(6.019$ & $1.864$ & $2.628) \times 10^{-7}$  & 12.99   & $(1.012$ & $0.013$ & $0.045) \times 10^{-1}$  & $(5.668$ & $0.086$ & $0.233) \times 10^{-3}$  \\
    708.6   & $(3.001$ & $0.965$ & $1.355) \times 10^{-7}$  & 13.82   & $(8.313$ & $0.104$ & $0.368) \times 10^{-2}$  & $(4.643$ & $0.071$ & $0.191) \times 10^{-3}$  \\
    892.9   & $(1.394$ & $0.513$ & $0.707) \times 10^{-7}$  & 14.69   & $(6.826$ & $0.086$ & $0.302) \times 10^{-2}$  & $(3.864$ & $0.060$ & $0.159) \times 10^{-3}$  \\
    1122.5  & $(6.370$ & $2.778$ & $3.739) \times 10^{-8}$  & 15.59   & $(5.642$ & $0.072$ & $0.250) \times 10^{-2}$  & $(3.262$ & $0.052$ & $0.134) \times 10^{-3}$  \\
    1414.5  & $(2.749$ & $1.316$ & $1.740) \times 10^{-8}$  & 16.52   & $(4.654$ & $0.060$ & $0.206) \times 10^{-2}$  & $(2.718$ & $0.045$ & $0.112) \times 10^{-3}$  \\
    1778.3  & $(1.095$ & $0.577$ & $0.736) \times 10^{-8}$  & 17.48   & $(3.888$ & $0.051$ & $0.173) \times 10^{-2}$  & $(2.293$ & $0.038$ & $0.094) \times 10^{-3}$  \\
    2240.8  & $(4.512$ & $2.501$ & $3.135) \times 10^{-9}$  & 18.48   & $(3.250$ & $0.043$ & $0.144) \times 10^{-2}$  & $(1.933$ & $0.032$ & $0.079) \times 10^{-3}$  \\
    2817.1  & $(1.997$ & $1.092$ & $1.376) \times 10^{-9}$  & 19.51   & $(2.740$ & $0.036$ & $0.122) \times 10^{-2}$  & $(1.666$ & $0.029$ & $0.068) \times 10^{-3}$  \\
    3549.7  & $(7.971$ & $4.720$ & $5.783) \times 10^{-10}$ & 20.58   & $(2.292$ & $0.031$ & $0.102) \times 10^{-2}$  & $(1.454$ & $0.025$ & $0.059) \times 10^{-3}$  \\
    4472.9  & $(3.520$ & $2.054$ & $2.529) \times 10^{-10}$ & 21.68   & $(1.925$ & $0.025$ & $0.086) \times 10^{-2}$  & $(1.214$ & $0.022$ & $0.050) \times 10^{-3}$  \\
    5636.2  & $(1.650$ & $0.879$ & $1.118) \times 10^{-10}$ & 22.83   & $(1.626$ & $0.022$ & $0.072) \times 10^{-2}$  & $(1.018$ & $0.018$ & $0.042) \times 10^{-3}$  \\
    7085.9  & $(7.103$ & $3.821$ & $4.844) \times 10^{-11}$ & 24.01   & $(1.381$ & $0.018$ & $0.061) \times 10^{-2}$  & $(9.031$ & $0.169$ & $0.369) \times 10^{-4}$  \\
    8928.7  & $(2.848$ & $1.638$ & $2.027) \times 10^{-11}$ & 25.25   & $(1.176$ & $0.016$ & $0.053) \times 10^{-2}$  & $(7.647$ & $0.146$ & $0.313) \times 10^{-4}$  \\
    11225.0 & $(1.515$ & $0.714$ & $0.955) \times 10^{-11}$ & 26.56   & $(9.911$ & $0.136$ & $0.442) \times 10^{-3}$  & $(6.757$ & $0.131$ & $0.276) \times 10^{-4}$  \\
    14145.0 & $(5.341$ & $3.076$ & $3.805) \times 10^{-12}$ & 27.95   & $(8.416$ & $0.116$ & $0.376) \times 10^{-3}$  & $(5.747$ & $0.115$ & $0.235) \times 10^{-4}$  \\
     &      &  &                                            & 29.43   & $(7.128$ & $0.098$ & $0.318) \times 10^{-3}$  & $(5.063$ & $0.102$ & $0.207) \times 10^{-4}$  \\
     &      &  &                                            & 31.00   & $(5.957$ & $0.083$ & $0.266) \times 10^{-3}$  & $(4.273$ & $0.089$ & $0.175) \times 10^{-4}$  \\
     &      &  &                                            & 32.66   & $(5.053$ & $0.071$ & $0.226) \times 10^{-3}$  & $(3.681$ & $0.079$ & $0.152) \times 10^{-4}$  \\
     &      &  &                                            & 34.43   & $(4.260$ & $0.061$ & $0.191) \times 10^{-3}$  & $(3.126$ & $0.069$ & $0.129) \times 10^{-4}$  \\
     &      &  &                                            & 36.32   & $(3.587$ & $0.052$ & $0.161) \times 10^{-3}$  & $(2.754$ & $0.062$ & $0.114) \times 10^{-4}$  \\
     &      &  &                                            & 38.33   & $(3.039$ & $0.043$ & $0.136) \times 10^{-3}$  & $(2.328$ & $0.055$ & $0.097) \times 10^{-4}$  \\
     &      &  &                                            & 40.48   & $(2.543$ & $0.037$ & $0.114) \times 10^{-3}$  & $(2.004$ & $0.048$ & $0.084) \times 10^{-4}$  \\
     &      &  &                                            & 42.78   & $(2.123$ & $0.031$ & $0.094) \times 10^{-3}$  & $(1.723$ & $0.043$ & $0.073) \times 10^{-4}$  \\
     &      &  &                                            & 45.26   & $(1.779$ & $0.026$ & $0.079) \times 10^{-3}$  & $(1.446$ & $0.037$ & $0.062) \times 10^{-4}$  \\
     &      &  &                                            & 47.92   & $(1.499$ & $0.023$ & $0.066) \times 10^{-3}$  & $(1.323$ & $0.034$ & $0.057) \times 10^{-4}$  \\
     &      &  &                                            & 50.80   & $(1.222$ & $0.019$ & $0.054) \times 10^{-3}$  & $(1.029$ & $0.029$ & $0.045) \times 10^{-4}$  \\
     &      &  &                                            & 53.92   & $(1.018$ & $0.016$ & $0.045) \times 10^{-3}$  & $(8.860$ & $0.254$ & $0.392) \times 10^{-5}$  \\
     &      &  &                                            & 57.32   & $(8.419$ & $0.130$ & $0.367) \times 10^{-4}$  & $(7.558$ & $0.223$ & $0.337) \times 10^{-5}$  \\
     &      &  &                                            & 61.03   & $(6.803$ & $0.107$ & $0.297) \times 10^{-4}$  & $(6.115$ & $0.191$ & $0.279) \times 10^{-5}$  \\
     &      &  &                                            & 65.11   & $(5.622$ & $0.090$ & $0.245) \times 10^{-4}$  & $(5.502$ & $0.173$ & $0.252) \times 10^{-5}$  \\
     &      &  &                                            & 69.62   & $(4.511$ & $0.073$ & $0.197) \times 10^{-4}$  & $(4.367$ & $0.145$ & $0.205) \times 10^{-5}$  \\
     &      &  &                                            & 74.65   & $(3.678$ & $0.060$ & $0.161) \times 10^{-4}$  & $(3.826$ & $0.127$ & $0.180) \times 10^{-5}$  \\
     &      &  &                                            & 80.29   & $(2.914$ & $0.048$ & $0.128) \times 10^{-4}$  & $(3.013$ & $0.106$ & $0.146) \times 10^{-5}$  \\
     &      &  &                                            & 86.69   & $(2.299$ & $0.039$ & $0.101) \times 10^{-4}$  & $(2.511$ & $0.091$ & $0.124) \times 10^{-5}$  \\
     &      &  &                                            & 94.02   & $(1.782$ & $0.031$ & $0.079) \times 10^{-4}$  & $(2.037$ & $0.076$ & $0.102) \times 10^{-5}$  \\
     &      &  &                                            & 102.60  & $(1.360$ & $0.024$ & $0.060) \times 10^{-4}$  & $(1.461$ & $0.064$ & $0.080) \times 10^{-5}$  \\
     &      &  &                                            & 112.70  & $(1.006$ & $0.018$ & $0.045) \times 10^{-4}$  & $(1.173$ & $0.052$ & $0.065) \times 10^{-5}$  \\
     &      &  &                                            & 125.00  & $(7.328$ & $0.139$ & $0.329) \times 10^{-5}$  & $(8.677$ & $0.405$ & $0.499) \times 10^{-6}$  \\
     &      &  &                                            & 140.10  & $(5.231$ & $0.105$ & $0.237) \times 10^{-5}$  & $(6.998$ & $0.328$ & $0.405) \times 10^{-6}$  \\
     &      &  &                                            & 158.90  & $(3.469$ & $0.073$ & $0.159) \times 10^{-5}$  & $(4.595$ & $0.236$ & $0.284) \times 10^{-6}$  \\
     &      &  &                                            & 183.10  & $(2.173$ & $0.049$ & $0.101) \times 10^{-5}$  & $(3.201$ & $0.174$ & $0.206) \times 10^{-6}$  \\
     &      &  &                                            & 216.20  & $(1.263$ & $0.032$ & $0.061) \times 10^{-5}$  & $(1.871$ & $0.118$ & $0.135) \times 10^{-6}$  \\
     &      &  &                                            & 261.80  & $(7.136$ & $0.206$ & $0.359) \times 10^{-6}$  & $(1.158$ & $0.081$ & $0.092) \times 10^{-6}$  \\
     &      &  &                                            & 326.80  & $(3.706$ & $0.126$ & $0.199) \times 10^{-6}$  & $(5.773$ & $0.518$ & $0.564) \times 10^{-7}$  \\
     &      &  &                                            & 428.50  & $(1.527$ & $0.066$ & $0.092) \times 10^{-6}$  & $(2.491$ & $0.300$ & $0.318) \times 10^{-7}$  \\
     &      &  &                                            & 588.80  & $(5.391$ & $0.351$ & $0.424) \times 10^{-7}$  & $(8.312$ & $1.843$ & $1.885) \times 10^{-8}$  \\
     &      &  &                                            & 832.30  & $(1.963$ & $0.208$ & $0.228) \times 10^{-7}$  & $(1.927$ & $1.174$ & $1.179) \times 10^{-8}$  \\
    \hline \hline
  \end{tabular}
  \caption{The CRE spectra of H.E.S.S. \cite{HESSICRC17} and 
  AMS-02 \cite{Aguilar:2019ksn}, and the CR positron spectrum of 
  AMS-02 \cite{Aguilar:2019owu}. The parameters $\sigma_{\mathrm{tot}}^{\prime}$
  and $\sigma_{\mathrm{tot}}$ represent the total uncertainties (quadratic sum
  of the statistical and systematic uncertainties) with and without including 
  the energy scale uncertainties, respectively. The flux $\Phi_{e^+ + e^-}$ 
  and $\Phi_{e^+}$ are in units of 
  $\mathrm{m}^{-2} \mathrm{sr}^{-1} \mathrm{s}^{-1} \mathrm{GeV}^{-1}$.}
  \label{tab:ESU_AMS}
  \end{table}

\clearpage
\section{Fitting results of models A to F}\label{app:para}

In this section, we summarize the fit results of each model
discussed in \Sec{sec:ModelAndResults}.
The prior ranges, best-fit values, statistical means 
and variations of the parameters of each model are summarized in 
\cref{tab:singlePWN,tab:catalogPWN,tab:singlePWNDM,tab:catalogPWNDM,tab:singlePWNSNR,tab:catalogPWNDMSNR}.
The Bayesian evidence and $\chi^2$ values of each fit are summarized in 
\Tab{tab:chi2}.

\begin{table}[h]
  \begin{tabular}{l R{1.3cm} @{ $\sim$ } L{1.3cm} C{2.4cm} C{2.4cm} C{2.4cm}}
    \hline \hline
                                                  Parameters & \multicolumn{2}{c}{Prior ranges} &  Best-fit &     Mean & $\sigma$ \\
  \hline
  $                                              \gamma_{e}$ &          2.0 &         3.5 &    2.621 &    2.617 &    0.009 \\
  $                                                   N_{e}$ &   $10^{-10}$ &   $10^{-8}$ & $1.146\times 10^{-9}$ & $1.139\times 10^{-9}$ & $1.246\times 10^{-11}$ \\
  $                                    \log(\rho_c/\rm MeV)$ &            5 &           8 &    7.917 &    7.733 &    0.186 \\
  $                          \log(T_{\mathrm{psr}}/\rm yrs)$ &            4 &           6 &    4.852 &    4.912 &    0.121 \\
  $                                 d_{\mathrm{psr}}/\rm pc$ &           80 &        1000 &      116 &      227 &    99 \\
  $                                   \gamma_{\mathrm{psr}}$ &          1.5 &         2.4 &    2.106 &    2.192 &    0.077 \\
  $                       \log(E_{c, \mathrm{psr}}/\rm GeV)$ &            2 &           5 &    3.501 &    3.840 &    0.392 \\
  $\log(\eta_{\mathrm{psr}}\dot{\mathcal{E}}_{\mathrm{psr}}/\rm erg~ s^{-1})$ &   32 &          37 &   35.129 &   35.334 &    0.212 \\
  \hline \hline
  \end{tabular}
  \caption{Parameters of \textit{Model-A} described in 
  \Sec{sec:c_sym_sources}, determined through fitting to the CRE and CR positron data. The 
  prior ranges, best-fit values, statistic means, and statistic variations are listed. 
  $N_e$ represents the post-propagated normalization flux of the primary electrons 
  at 25 GeV, which is in units of $\rm cm^{-2} sr^{-1} s^{-1} MeV^{-1}$.}
  \label{tab:singlePWN}
\end{table}

\begin{table}[h]
  \begin{tabular}{L{2.5cm} R{1.3cm} @{ $\sim$ } L{1.3cm} C{2.4cm} C{2.4cm} C{2.4cm}}
  \hline \hline
               Parameters & \multicolumn{2}{c}{Prior ranges} & Best-fit & Mean & $\sigma$ \\
  \hline
  $             \gamma_e$ &          2.0 &         3.5 &    2.606 &    2.603 &    0.009 \\
  $                  N_e$ &   $10^{-10}$ &   $10^{-8}$ & $1.128\times 10^{-9}$ & $1.127\times 10^{-9}$ & $1.274\times 10^{-11}$ \\
  $ \log(\rho_c/\rm MeV)$ &            5 &           8 &    7.978 &    7.713 &    0.190 \\
  $                 \eta$ &            0 &           1 &    0.097 &    0.099 &    0.005 \\
  $               \gamma$ &          1.5 &         2.4 &    1.792 &    1.799 &    0.023 \\
  $    \log(E_c/\rm GeV)$ &            2 &           5 &    3.933 &    3.988 &    0.133 \\
  \hline \hline
  \end{tabular}
  \caption{Parameters of \textit{Model-B} described in 
  \Sec{sec:c_sym_sources}, determined through fitting to the CRE and CR positron data. The 
  prior ranges, best-fit values, statistic means, and statistic variations are listed. 
  $N_e$ represents the post-propagated normalization flux of the primary electrons 
  at 25 GeV, which is in units of $\rm cm^{-2} sr^{-1} s^{-1} MeV^{-1}$.}
  \label{tab:catalogPWN}
\end{table}

\begin{table}[h]
  \begin{tabular}{l R{1.3cm} @{ $\sim$ } L{1.3cm} C{2.4cm} C{2.4cm} C{2.4cm}}
    \hline \hline
                                                  Parameters & \multicolumn{2}{c}{Prior ranges} &  Best-fit &     Mean & $\sigma$ \\
  \hline
  $                                              \gamma_{e}$ &          2.0 &         3.5 &    2.624 &    2.617 &    0.009 \\
  $                                                   N_{e}$ &   $10^{-10}$ &   $10^{-8}$ & $1.149\times 10^{-9}$ & $1.140\times 10^{-9}$ & $1.252\times 10^{-11}$ \\
  $                                    \log(\rho_c/\rm MeV)$ &            5 &           8 &    7.987 &    7.737 &    0.182 \\
  $                          \log(T_{\mathrm{psr}}/\rm yrs)$ &            4 &           6 &    4.763 &    4.916 &    0.124 \\
  $                                 d_{\mathrm{psr}}/\rm pc$ &           80 &        1000 &      149 &      231 &    100 \\
  $                                   \gamma_{\mathrm{psr}}$ &          1.5 &         2.4 &    2.097 &    2.198 &    0.077 \\
  $                       \log(E_{c, \mathrm{psr}}/\rm GeV)$ &            2 &           5 &    3.368 &    3.859 &    0.405 \\
  $\log(\eta_{\mathrm{psr}}\dot{\mathcal{E}}_{\mathrm{psr}}/\rm erg~ s^{-1})$ &   32 &          37 &   35.181 &   35.339 &    0.212 \\
  $                              \log(m_{\rm \chi}/\rm GeV)$ &            1 &           4 &    1.663 &    2.569 &    1.086 \\
  $       \log(\langle{\sigma v}\rangle/(\rm cm^{3}s^{-1}))$ &          -26 &         -21 &  -25.988 &  -24.557 &    1.085 \\
  \hline \hline
  \end{tabular}
  \caption{Parameters of \textit{Model-C}  described in 
  \Sec{sec:c_sym_sources}, determined through fitting to the CRE and CR positron data. The 
  prior ranges, best-fit values, statistic means, and statistic variations are listed. 
  $N_e$ represents the post-propagated normalization flux of the primary electrons 
  at 25 GeV, which is in units of $\rm cm^{-2} sr^{-1} s^{-1} MeV^{-1}$.}
  \label{tab:singlePWNDM}
\end{table}

\begin{table}[h]
  \begin{tabular}{l R{1.3cm} @{ $\sim$ } L{1.3cm} C{2.4cm} C{2.4cm} C{2.4cm}}
  \hline \hline
               Parameters & \multicolumn{2}{c}{Prior ranges} & Best-fit & Mean & $\sigma$ \\
  \hline
  $                                         \gamma_e$ &          2.0 &         3.5 &    2.623 &    2.617 &    0.009  \\
  $                                              N_e$ &   $10^{-10}$ &   $10^{-8}$ & $1.142\times 10^{-9}$ & $1.140\times 10^{-9}$ & $1.217\times 10^{-11}$ \\
  $                             \log(\rho_c/\rm MeV)$ &            5 &           8 &    7.974 &    7.744 &    0.179 \\
  $                                             \eta$ &            0 &           1 &    0.097 &    0.093 &    0.009 \\
  $                                           \gamma$ &          1.5 &         2.4 &    2.038 &    2.033 &    0.055 \\
  $                                \log(E_c/\rm GeV)$ &            2 &           5 &    4.891 &    4.457 &    0.471 \\
  $                       \log(m_{\rm \chi}/\rm GeV)$ &            1 &           4 &    3.269 &    3.281 &    0.095 \\
  $\log(\langle{\sigma v}\rangle/(\rm cm^{3}s^{-1}))$ &          -26 &         -21 &  -23.302 &  -23.298 &    0.159 \\
  \hline \hline
  \end{tabular}
  \caption{Parameters of \textit{Model-D} described in 
  \Sec{sec:c_sym_sources}, determined through fitting to the CRE and CR positron data. The 
  prior ranges, best-fit values, statistic means, and statistic variations are listed. 
  $N_e$ represents the post-propagated normalization flux of the primary electrons 
  at 25 GeV, which is in units of $\rm cm^{-2} sr^{-1} s^{-1} MeV^{-1}$.}
  \label{tab:catalogPWNDM}
\end{table}

\begin{table}[h]
  \begin{tabular}{l R{1.3cm} @{ $\sim$ } L{1.3cm} C{2.4cm} C{2.4cm} C{2.4cm}}
    \hline \hline
                                                  Parameters & \multicolumn{2}{c}{Prior ranges} &  Best-fit &     Mean & $\sigma$ \\
  \hline
  $                                              \gamma_{e}$ &          2.0 &         3.5 &    2.706 &    2.671 &    0.025 \\
  $                                                   N_{e}$ &   $10^{-10}$ &   $10^{-8}$ & $1.180\times 10^{-9}$ & $1.159\times 10^{-9}$ & $1.445\times 10^{-11}$ \\
  $                                    \log(\rho_c/\rm MeV)$ &            5 &           8 &    7.996 &    7.756 &    0.167 \\
  $                          \log(T_{\mathrm{psr}}/\rm yrs)$ &            4 &           6 &    4.721 &    4.829 &    0.204 \\
  $                                d_{\mathrm{psr}}/\rm pc $ &           80 &        1000 &      134 &      229 &    109 \\
  $                                   \gamma_{\mathrm{psr}}$ &          1.5 &         2.4 &    2.052 &    2.166 &    0.096 \\
  $                       \log(E_{c, \mathrm{psr}}/\rm GeV)$ &            2 &           5 &    2.905 &    3.254 &    0.361 \\
  $\log(\eta_{\mathrm{psr}}\dot{\mathcal{E}}_{\mathrm{psr}}/\rm erg~ s^{-1})$ &   32 &          37 &   35.139 &   35.346 &    0.232 \\
  $                          \log(T_{\mathrm{snr}}/\rm yrs)$ &            4 &           6 &    4.754 &    4.771 &    0.245 \\
  $                                d_{\mathrm{snr}}/\rm pc $ &           80 &        2000 &      523 &      466 &      258 \\
  $                                   \gamma_{\mathrm{snr}}$ &          1.5 &         2.6 &    2.166 &    1.926 &    0.200 \\
  $                       \log(E_{c, \mathrm{snr}}/\rm GeV)$ &            2 &           5 &    3.613 &    3.571 &    0.433 \\
  $                     \log(E_{\mathrm{tot, snr}}/\rm erg)$ &           45 &          49 &   48.743 &   48.034 &    0.591 \\
  \hline \hline
  \end{tabular}
  \caption{Parameters of \textit{Model-E}  described in 
  \Sec{sec:c_asym_sources}, determined through fitting to the CRE and CR positron data. The 
  prior ranges, best-fit values, statistic means, and statistic variations are listed. 
  $N_e$ represents the post-propagated normalization flux of the primary electrons 
  at 25 GeV, which is in units of $\rm cm^{-2} sr^{-1} s^{-1} MeV^{-1}$.}
  \label{tab:singlePWNSNR}
\end{table}

\begin{table}[h]
  \begin{tabular}{l R{1.3cm} @{ $\sim$ } L{1.3cm} C{2.4cm} C{2.4cm} C{2.4cm}}
  \hline \hline
               Parameters & \multicolumn{2}{c}{Prior ranges} & Best-fit & Mean & $\sigma$ \\
  \hline
  $                                         \gamma_e$ &          2.0 &         3.5 &    2.675 &    2.659 &    0.023  \\
  $                                              N_e$ &   $10^{-10}$ &   $10^{-8}$ & $1.157\times 10^{-9}$ & $1.159\times 10^{-9}$ & $1.459\times 10^{-11}$ \\
  $                             \log(\rho_c/\rm MeV)$ &            5 &           8 &    7.924 &    7.754 &    0.168 \\
  $                                             \eta$ &            0 &           1 &    0.098 &    0.091 &    0.008 \\
  $                                           \gamma$ &          1.5 &         2.4 &    1.995 &    2.039 &    0.055 \\
  $                                \log(E_c/\rm GeV)$ &            2 &           5 &    4.915 &    4.485 &    0.434 \\
  $                       \log(m_{\rm \chi}/\rm GeV)$ &            1 &           4 &    3.032 &    3.052 &    0.129 \\
  $\log(\langle{\sigma v}\rangle/(\rm cm^{3}s^{-1}))$ &          -26 &         -21 &  -23.771 &  -23.669 &    0.213 \\
  $                   \log(T_{\mathrm{snr}}/\rm yrs)$ &            4 &           6 &    5.182 &    4.953 &    0.255 \\
  $                         d_{\mathrm{snr}}/\rm pc $ &           80 &        2000 &      543 &      622 &      315 \\
  $                            \gamma_{\mathrm{snr}}$ &          1.5 &         2.6 &    1.903 &    1.812 &    0.211 \\
  $                \log(E_{c, \mathrm{snr}}/\rm GeV)$ &            2 &           5 &    4.980 &    3.618 &    0.483 \\
  $              \log(E_{\mathrm{tot, snr}}/\rm erg)$ &           45 &          49 &   48.615 &   48.129 &    0.683 \\
  \hline \hline
  \end{tabular}
  \caption{Parameters of \textit{Model-F} described in 
  \Sec{sec:c_asym_sources}, determined through fitting to the CRE and CR positron data. The 
  prior ranges, best-fit values, statistic means, and statistic variations are listed. 
  $N_e$ represents the post-propagated normalization flux of the primary electrons 
  at 25 GeV, which is in units of $\rm cm^{-2} sr^{-1} s^{-1} MeV^{-1}$.}
  \label{tab:catalogPWNDMSNR}
\end{table}

\begin{table}[h]
  \begin{tabular}{l C{2.6cm} C{1.4cm} C{1.2cm} C{3.0cm} } \hline \hline
  Models & $\chi^2_\text{tot}$/d.o.f & $\chi^2_{e^+ + e^-}$ & $\chi^2_{e^+}$ & log-evidence \\  \hline
  \textit{Model-A (single PWN):}                & 212.8/188 &         190.5 & 22.3 & 2389.7 $\pm$ 0.2 \\
  \textit{Model-B (multiple PWNe):}             & 263.3/190 &         228.2 & 35.1 & 2367.0 $\pm$ 0.2 \\  
  \hline
  \textit{Model-C (single PWN + DM):}           & 212.6/186 &         187.6 & 25.0 & 2389.9 $\pm$ 0.2 \\
  \textit{Model-D (multiple PWNe + DM):}        & 217.7/188 &         189.3 & 28.4 & 2385.0 $\pm$ 0.2 \\  
  \hline
  \textit{Model-E (single PWN + SNR):}          & 188.9/183 &         178.1 & 10.8 & 2394.1 $\pm$ 0.2 \\
  \textit{Model-F (multiple PWNe + DM + SNR):}  & 192.3/183 &         177.1 & 15.2 & 2388.1 $\pm$ 0.2 \\  
  \hline \hline
  \end{tabular}
  \caption{Values of the fitting $\chi^2$ and the logarithmic Bayesian evidence
  of each model. The number of data points for the CRE and CR positron are 161 
  and 35, respectively.}
  \label{tab:chi2}
\end{table}

\end{document}